\documentclass[12pt,twoside, a4paper]{article}
\def\pd{\partial}
\def\mc{\mathcal}
\def\ul{\underline}

\usepackage[dvips]{graphicx}
\usepackage{amssymb}
\usepackage[bbgreekl]{mathbbol}
\usepackage{amssymb,amsmath}
\usepackage{graphicx}
\usepackage{dsfont}
\usepackage{caption}
\usepackage{subcaption}
\usepackage{mathtools}
\usepackage{verbatim}
\usepackage{graphicx}
\usepackage{multirow}
\usepackage[outline]{contour}
\usepackage{xcolor,colortbl}
\input{epsf.sty}  
\begin{document}
\begin{center}
\Large{\textbf{Supersymmetric $AdS_6$ black holes from ISO(3)$\times$U(1) $F(4)$ gauged supergravity}}
\end{center}
\vspace{1 cm}
\begin{center}
\large{\textbf{Parinya Karndumri}}
\end{center}
\begin{center}
String Theory and Supergravity Group, Department
of Physics, Faculty of Science, Chulalongkorn University, 254 Phayathai Road, Pathumwan, Bangkok 10330, Thailand \\
E-mail: parinya.ka@hotmail.com \vspace{1 cm}
\end{center}
\begin{abstract}
We study supersymmetric $AdS_6$ black holes from matter-coupled $F(4)$ gauged supergravity coupled to four vector multiplets with $ISO(3)\times U(1)$ gauge group. This gauged supergravity admits a maximally supersymmetric $AdS_6$ vacuum with $SO(3)\subset ISO(3)$ symmetry. We find a number of new supersymmetric $AdS_2\times \mc{M}_4$ solutions by performing topological twists along $\mc{M}_4$. For $\mc{M}_4$ being a product of two Riemann surfaces $\Sigma\times \widetilde{\Sigma}$, we perform a twist by $SO(2)\times U(1)$ gauge fields and find $AdS_2\times \Sigma\times \widetilde{\Sigma}$ solutions for at least one of the Riemann surface being negatively curved. For $\mc{M}_4$ being a Kahler four-cycle $\mc{M}_{\textrm{K}4}$, a twist by $SO(2)\times U(1)$ gauge fields leads to $AdS_2\times \mc{M}_{\textrm{K}4}^-$ solutions for negatively curved $\mc{M}_{\textrm{K}4}$. Finally, performing a twist by turning on $SO(3)$ gauge fields in the case of $\mc{M}_4$ being a Cayley four-cycle $\mc{M}_{\textrm{C}4}$ also leads to $AdS_2\times \mc{M}_{\textrm{C}4}^-$ solutions for negatively curved $\mc{M}_{\textrm{C}4}$. We give numerical black hole solutions interpolating between all of these $AdS_2\times \mc{M}_4$ near horizon geometries and the asymptotically locally $AdS_6$ vacuum. It is also possible to uplift all of these solutions to type IIB theory via consistent truncations on $S^2\times \Sigma$ leading to a new class of supersymmetric $AdS_2\times \mc{M}_4\times S^2\times \Sigma$ solutions.   
\end{abstract}
\newpage
\section{Introduction}
One of the most challenging problems in quantum gravity is to determine the microscopic origin of black hole entropy. Since the original result of \cite{SV_BH_entropy}, a large number of subsequent works has successfully derived the entropy of different types of asymptotically flat black holes by counting the number of the associated microstates. On the other hand, analogous results for black holes in asymptotically AdS spaces have appeared only recently. In this case, the Bekenstein-Hawking entropy can be holographically computed by using topologically twisted indices in the dual field theories \cite{Twisted_index1,Twisted_index2,Twisted_index3}, see also \cite{BH_counting1}-\cite{BH_counting13}. This is another great achievement of the AdS/CFT correspondence \cite{maldacena,Gubser_AdS_CFT,Witten_AdS_CFT} in providing more insights into various aspects of strongly-coupled field theories. This framework has also been extended to the case of black strings in AdS spaces \cite{black_string_entropy,black_string_entropy1,black_string_entropy2} and is conjectured to be applicable to other black branes of different dimensionalities.
\\
\indent The aforementioned results have been worked out mostly for supersymmetric AdS black holes in which both gravity and field theory sides are well under control. Accordingly, finding new supersymmetric black hole solutions in asymptotically AdS spaces is essential to test various aspects of the proposed formalism and to gain more insights into the quantum dynamics of AdS black holes. In this paper, we will consider supersymmetric black hole solutions in asymptotically $AdS_6$ spaces. It is well-known that only the half-maximal $N=(1,1)$ or $F(4)$ gauged supergravity can lead to supersymmetric $AdS_6$ vacua. The pure $F(4)$ gauged supergravity with $SU(2)$ gauge group has been first constructed in \cite{F4_Romans}, and the extension to $F(4)$ gauged supergravity coupled to $n$ vector multiplets with a compact $SU(2)\times G_c$ gauge group has been constructed in \cite{F4SUGRA1,F4SUGRA2}. More recently, general forms of gauge groups that lead to supersymmetric $AdS_6$ vacua have also been classified in \cite{AdS6_Jan}. 
\\
\indent A number of $AdS_6$ black hole solutions with near horizon geometries of the form $AdS_2\times \mc{M}_4$ from pure $F(4)$ gauged supergravity has been considered in \cite{Naka} and more recently in \cite{AdS6_BH_Minwoo1,flow_across_bobev}. Solutions in the matter-coupled $F(4)$ gauged supergravity with $SU(2)\times U(1)$ and $SU(2)\times SU(2)$ gauge groups have subsequently appeared in \cite{AdS6_BH_Zaffaroni,AdS6_BH_Minwoo,AdS6_BH}. The pure $F(4)$ gauged supergravity and the matter-coupled theory with $SU(2)\times U(1)$ gauge group can be embedded respectively in massive type IIA and type IIB theories \cite{Massive_IIA_onS4,Henning_Malek_AdS7_6}. On the other hand, the matter-coupled $F(4)$ gauged supergravity with $SU(2)\times SU(2)$ gauge group admits a much richer class of supersymmetric $AdS_6$ black holes but currently has no known higher-dimensional origins.        
\\
\indent In this work, we look for more interesting supersymmetric $AdS_6$ black hole solutions within $F(4)$ gauged supergravity coupled to four vector multiplets with $ISO(3)\times U(1)$ gauge group. As shown in \cite{Henning_Malek_AdS7_6}, this gauged supergravity might arise from a consistent truncation of type IIB theory on $S^2\times \Sigma$ with $\Sigma$ being a Riemann surface. The matter-coupled supergravity in this case has $\mathbb{R}^+\times SO(4,4)$ global symmetry, and the $ISO(3)$ part of the gauge group is embedded in $SO(4,3)\subset SO(4,4)$. The $U(1)$ factor on the other hand corresponds to the abelian gauge symmetry of the fourth vector multiplet as in the ungauged supergravity. A number of holographic solutions from this gauged supergravity has appeared recently in \cite{ISO3_flow} and \cite{ISO3_defect}. In the present work, we will look for supersymmetric $AdS_6$ black hole solutions from this gauged supergravity by repeating the analysis of \cite{AdS6_BH} in the case of compact $SU(2)\times SU(2)$ gauge group. This would provide more examples of known $AdS_6$ black hole solutions and could possibly lead to a new class of $AdS_6$ black holes in type IIB theory upon uplifting to ten dimensions.   
\\
\indent The paper is organized as follows. We give a brief review of the matter-coupled $F(4)$ gauged supergravity with $ISO(3)\times U(1)$ gauge group in section \ref{6D_SO4gaugedN2} together with the supersymmetric $AdS_6$ vacuum preserving $SO(3)\subset ISO(3)$ symmetry. By performing topological twists using $SO(2)\times U(1)$ gauge fields, we study $AdS_2\times \Sigma\times\widetilde{\Sigma}$ solutions and find a new class of $AdS_2\times H^2\times H^2$, $AdS_2\times S^2\times H^2$ and $AdS_2\times H^2\times S^2$ solutions in section \ref{AdS2_Sigma_Sigma}. Examples of numerical black hole solutions interpolating between the $AdS_6$ vacuum and these near horizon geometries are also given. We then perform a similar analysis for the case of $AdS_2\times \mc{M}_4$ solutions with $\mc{M}_4$ given by a Kahler four-cycle and a Cayley four-cycle in section \ref{AdS2_M4} and find $AdS_6$ black hole solutions with horizon geometries given by negatively curved Kahler and Cayley four-cycles. We end the paper by giving some conclusions and comments in section \ref{conclusion}. 
\section{Matter-coupled $F(4)$ gauged supergravity in six dimensions}\label{6D_SO4gaugedN2}
We first review the general structure of matter-coupled $F(4)$ gauged supergravity in six dimensions. All the conventions are mostly the same as \cite{F4SUGRA1,F4SUGRA2} to which we refer for more detail. In this paper, we will mainly present relevant formulae for obtaining supersymmetric $AdS_6$ black hole solutions. The field content of the supergravity coupled to $n$ vector multiplets is collectively given by 
\begin{equation}
\left(e^{\hat{\mu}}_\mu,\psi^A_\mu, A^\Lambda_\mu, B_{\mu\nu}, \chi^A,\lambda^I_A,\phi^{I\alpha}\right) .
\end{equation}
The metric signature is $(-+++++)$ with space-time and tangent space indices denoted respectively by $\mu,\nu=0,\ldots ,5$ and $\hat{\mu},\hat{\nu}=0,\ldots, 5$. The bosonic fields are given by the graviton $e^{\hat{\mu}}_\mu$, a two-form field $B_{\mu\nu}$, $4+n$ vector fields $A^\Lambda=(A^\alpha_\mu,A^I)$, the dilaton $\sigma$ and $4n$ scalars $\phi^{\alpha I}$. The fermionic fields are given by two gravitini $\psi^A_\mu$, two spin-$\frac{1}{2}$ fields $\chi^A$ and $2n$ gaugini $\lambda^I_A$. Indices $\alpha=(0,r)=0,1,2,3$ are split into $SU(2)_R\sim USp(2)_R\sim SO(3)_R$ R-symmetry singlet $(0)$ and adjoint $(r=1,2,3)$ while $I=1,2,\ldots, n$ denote the $n$ vector multiplets. The $SU(2)_R$ fundamental indices $A,B,\ldots =1,2$ are raised and lowered by $\epsilon^{AB}$ and $\epsilon_{AB}$ with the convention $T^A=\epsilon^{AB}T_B$ and $T_A=T^B\epsilon_{BA}$. 
\\
\indent The dilaton and $4n$ scalar fields from the vector multiplets are described by $\mathbb{R}^+\times SO(4,n)/SO(4)\times SO(n)$ coset with $\mathbb{R}^+$ corresponding to the dilaton. The $SO(4,n)/SO(4)\times SO(n)$ can be parametrized by a coset
representative ${L^\Lambda}_{\ul{\Sigma}}$ transforming under the global $SO(4,n)$ and local $SO(4)\times SO(n)$ symmetries by left and right multiplications respectively with indices $\Lambda,\ul{\Sigma}=0,\ldots , n+3$. The local index can be decomposed into $\ul{\Sigma}=(\alpha,I)=(0,r,I)$ leading to the following components of the coset representative 
\begin{equation}
{L^\Lambda}_{\ul{\Sigma}}=({L^\Lambda}_\alpha,{L^\Lambda}_I).
\end{equation}
The inverse of ${L^\Lambda}_{\ul{\Sigma}}$ will be denoted by ${(L^{-1})^{\ul{\Lambda}}}_\Sigma=({(L^{-1})^{\alpha}}_\Sigma,{(L^{-1})^{I}}_\Sigma)$. $SO(4,n)$ indices are raised and lowered by the invariant tensor
\begin{equation}
\eta_{\Lambda\Sigma}=\eta^{\Lambda\Sigma}=(\delta_{\alpha\beta},-\delta_{IJ}).
\end{equation}
\indent The bosonic Lagrangian of the matter-coupled gauged supergravity can be written as
\begin{eqnarray}
e^{-1}\mathcal{L}&=&\frac{1}{4}R-e\pd_\mu \sigma\pd^\mu \sigma
-\frac{1}{4}P^{I\alpha}_\mu P^{\mu}_{I\alpha}-\frac{1}{8}e^{-2\sigma}\mc{N}_{\Lambda\Sigma}\widehat{F}^\Lambda_{\mu\nu}
\widehat{F}^{\Sigma\mu\nu}-\frac{3}{64}e^{4\sigma}H_{\mu\nu\rho}H^{\mu\nu\rho}\nonumber \\
& &-V-\frac{1}{64}e^{-1}\epsilon^{\mu\nu\rho\sigma\lambda\tau}B_{\mu\nu}\left(\eta_{\Lambda\Sigma}\widehat{F}^\Lambda_{\rho\sigma}
\widehat{F}^\Sigma_{\lambda\tau}+mB_{\rho\sigma}\widehat{F}^\Lambda_{\lambda\tau}\delta^{\Lambda 0}+\frac{1}{3}m^2B_{\rho\sigma}B_{\lambda\tau}\right)\nonumber \\
\label{Lar}
\end{eqnarray}
with $e=\sqrt{-g}$. The corresponding field strength tensors are defined by
\begin{eqnarray}
\widehat{F}^\Lambda=F^\Lambda-m\delta^{\Lambda 0}B, \qquad F^\Lambda=dA^\Lambda-\frac{1}{2}f^{\phantom{\Sigma}\Lambda}_{\Sigma\phantom{\Lambda}\Gamma} A^\Sigma\wedge A^\Gamma,\qquad H=dB\, .
\end{eqnarray}
$f^{\phantom{\Sigma}\Lambda}_{\Sigma\phantom{\Lambda}\Gamma}$ are structure constants of the gauge group. It is also useful to note the convention on differential forms used in \cite{F4SUGRA1, F4SUGRA2}
\begin{equation}
F^\Lambda=F^\Lambda_{\mu\nu} dx^\mu \wedge dx^\nu\qquad \textrm{and}\qquad H=H_{\mu\nu\rho}dx^\mu \wedge dx^\nu\wedge dx^\rho
\end{equation}
which is different from the usual convention and, in particular, leads to
\begin{eqnarray}
F_{\mu\nu}&=&\frac{1}{2}(\pd_\mu A_\nu-\pd_\nu A_\mu-f^{\phantom{\Sigma}\Lambda}_{\Sigma\phantom{\Lambda}\Gamma}A^\Sigma_\mu A^\Gamma_\nu)\nonumber \\
 \textrm{and}\qquad H_{\mu\nu\rho}&=&\frac{1}{3}(\pd_\mu B_{\nu\rho}+\pd_\nu B_{\rho\mu}+\pd_\rho B_{\mu\nu})
\end{eqnarray}
\indent The vielbein on $SO(4,n)/SO(4)\times SO(n)$, $P^{I\alpha}_\mu=P^{I\alpha}_x\pd_\mu\phi^x$, $x=1,\ldots, 4n$, appearing in the scalar kinetic term, can be obtained from the left-invariant 1-form
\begin{equation}
{\Omega^{\ul{\Lambda}}}_{\ul{\Sigma}}=
{(L^{-1})^{\ul{\Lambda}}}_{\Pi}\nabla {L^\Pi}_{\ul{\Sigma}}
\qquad \textrm{with}\qquad \nabla
{L^\Lambda}_{\ul{\Sigma}}={dL^\Lambda}_{\ul{\Sigma}}
-f^{\phantom{\Gamma}\Lambda}_{\Gamma\phantom{\Lambda}\Pi}A^\Gamma
{L^\Pi}_{\ul{\Sigma}}
\end{equation}
via  
\begin{equation}
{P^I}_{\alpha}=({P^I}_{0},{P^I}_{r})=(\Omega^I_{\phantom{a}0},\Omega^I_{\phantom{a}r}).
\end{equation}
The other components of ${\Omega^{\ul{\Lambda}}}_{\ul{\Sigma}}$ are identified as the $SO(4)\times SO(n)$ composite connections $(\Omega^{rs},\Omega^{r0},\Omega^{IJ})$. 
\\
\indent The symmetric scalar matrix $\mc{N}_{\Lambda\Sigma}$, appearing in the vector kinetic term, is defined by
\begin{eqnarray}
\mc{N}_{\Lambda\Sigma}=L_{\Lambda\alpha}{(L^{-1})^\alpha}_\Sigma-L_{\Lambda I}{(L^{-1})^I}_\Sigma=(\eta L L^T\eta)_{\Lambda\Sigma}\, .
\end{eqnarray}
The scalar potential is given by
\begin{eqnarray}
V&=&-e^{2\sigma}\left[\frac{1}{36}A^2+\frac{1}{4}B^iB_i+\frac{1}{4}\left(C^I_{\phantom{s}t}C_{It}+4D^I_{\phantom{s}t}D_{It}\right)\right]
+m^2e^{-6\sigma}\mc{N}_{00}\nonumber \\
& &-me^{-2\sigma}\left[\frac{2}{3}AL_{00}-2B^iL_{0i}\right]
\end{eqnarray}
with various components of fermion-shift matrices defined by 
\begin{eqnarray}
A&=&\epsilon^{rst}K_{rst},\qquad B^i=\epsilon^{ijk}K_{jk0},\\
C^{\phantom{ts}t}_I&=&\epsilon^{trs}K_{rIs},\qquad D_{It}=K_{0It}\, .
\end{eqnarray}
The ``boosted'' structure constants are in turn given by
\begin{eqnarray}
K_{rs\alpha}&=&f^{\phantom{\Lambda}\Gamma}_{\Lambda\phantom{\Gamma}\Sigma}{L^\Lambda}_r(L^{-1})_{s\Gamma}{L^\Sigma}_\alpha,\nonumber \\
K_{\alpha It}&=&f^{\phantom{\Lambda}\Gamma}_{\Lambda\phantom{\Gamma}\Sigma}{L^\Lambda}_\alpha (L^{-1})_{I\Gamma}{L^\Sigma}_t
\end{eqnarray}
for $\alpha=(0,r)$. 
\\ 
\indent Finally, supersymmetry transformations of fermionic fields are given by
\begin{eqnarray}
\delta\psi_{\mu
A}&=&D_\mu\epsilon_A-\frac{1}{24}\left(Ae^\sigma+6me^{-3\sigma}(L^{-1})_{00}\right)\epsilon_{AB}\gamma_\mu\epsilon^B\nonumber
\\
& &-\frac{1}{8}
\left(B_te^\sigma-2me^{-3\sigma}(L^{-1})_{t0}\right)\gamma^7\sigma^t_{AB}\gamma_\mu\epsilon^B\nonumber \\
&
&+\frac{i}{16}e^{-\sigma}\left[\epsilon_{AB}(L^{-1})_{0\Lambda}\gamma_7+\sigma^r_{AB}(L^{-1})_{r\Lambda}\right]
F^\Lambda_{\nu\lambda}(\gamma_\mu^{\phantom{s}\nu\lambda}
-6\delta^\nu_\mu\gamma^\lambda)\epsilon^B\nonumber \\
& &+\frac{i}{32}e^{2\sigma}H_{\nu\lambda\rho}\gamma_7({\gamma_\mu}^{\nu\lambda\rho}-3\delta_\mu^\nu\gamma^{\lambda\rho})\epsilon_A,\label{delta_psi}\\
\delta\chi_A&=&\frac{1}{2}\gamma^\mu\pd_\mu\sigma\epsilon_{AB}\epsilon^B+\frac{1}{24}
\left[Ae^\sigma-18me^{-3\sigma}(L^{-1})_{00}\right]\epsilon_{AB}\epsilon^B\nonumber
\\
& &-\frac{1}{8}
\left[B_te^\sigma+6me^{-3\sigma}(L^{-1})_{t0}\right]\gamma^7\sigma^t_{AB}\epsilon^B\nonumber
\\
& &-\frac{i}{16}e^{-\sigma}\left[\sigma^r_{AB}(L^{-1})_{r\Lambda}-\epsilon_{AB}(L^{-1})_{0\Lambda}\gamma_7\right]F^\Lambda_{\mu\nu}\gamma^{\mu\nu}\epsilon^B\nonumber \\
& &-\frac{i}{32}e^{2\sigma}H_{\nu\lambda\rho}\gamma_7\gamma^{\nu\lambda\rho}\epsilon_A,
\label{delta_chi}\\
\delta
\lambda^{I}_A&=&P^I_{ri}\gamma^\mu\pd_\mu\phi^i\sigma^{r}_{\phantom{s}AB}\epsilon^B+P^I_{0i}
\gamma^7\gamma^\mu\pd_\mu\phi^i\epsilon_{AB}\epsilon^B-\left(2i\gamma^7D^I_{\phantom{s}t}+C^I_{\phantom{s}t}\right)
e^\sigma\sigma^t_{AB}\epsilon^B \nonumber
\\
& &+2me^{-3\sigma}(L^{-1})^I_{\phantom{ss}0}
\gamma^7\epsilon_{AB}\epsilon^B-\frac{i}{2}e^{-\sigma}(L^{-1})^I_{\phantom{s}\Lambda}F^\Lambda_{\mu\nu}
\gamma^{\mu\nu}\epsilon_{A}\label{delta_lambda}
\end{eqnarray}
with ${\sigma^{rA}}_B$ being Pauli matrices. The covariant derivative of $\epsilon_A$ is defined by
\begin{equation}
D_\mu \epsilon_A=\pd_\mu
\epsilon_A+\frac{1}{4}\omega_\mu^{ab}\gamma_{ab}\epsilon_A+Q_{AB}\epsilon^B\, .
\end{equation}
in which the composite connection is defined by
\begin{equation}
Q_{AB}=\frac{i}{2}\sigma^r_{AB}
\left[\frac{1}{2}\epsilon^{rst}\Omega_{\mu st}-i\gamma_7
\Omega_{\mu r0}\right].
\end{equation}
\indent In this paper, we are interested in a particular case of $n=4$ vector multiplets leading to $\mathbb{R}^+\times SO(4,4)$ global symmetry. The gauge group of interest here is $ISO(3)\times U(1)$ with $ISO(3)$ embedded in $SO(4,3)\subset SO(4,4)$ and the $U(1)$ factor is the abelian gauge symmetry associated with the fourth vector multiplet. The structure constants corresponding to the $ISO(3)\sim SO(3)\ltimes \mathbf{R}^3$ factor are given by
\begin{eqnarray}
f_{rst}=g\epsilon_{rst},\qquad f_{r\bar{s}\bar{t}}=-g\epsilon_{r\bar{s}\bar{t}},\qquad f_{\bar{r}\bar{s}\bar{t}}=-2g\epsilon_{\bar{r}\bar{s}\bar{t}}
\end{eqnarray}
with indices $I,J=1,2,3,4$ split as $I=(\bar{r},4)$ and $\bar{r},\bar{s}=1,2,3$. The corresponding gauge coupling constant is denoted by $g$. The gauge generators are given by $X_\Lambda=(X_0,X_r,X_{\bar{r}})$ with $X_0=0$, and $X_r$ and $X_r-X_{\bar{r}}$ respectively generate the $SO(3)$ compact subgroup and three abelian translations $\mathbf{R}^3$ transforming as adjoint representation of $SO(3)$, see more detail in \cite{ISO3_flow}.
\\
\indent We end this section by pointing out that this gauged supergravity admits a supersymmetric $AdS_6$ vacuum at the origin of the $SO(4,4)/SO(4)\times SO(4)$ scalar manifold with
\begin{eqnarray}
& &{L^\Lambda}_{\ul{\Sigma}}=\delta^\Lambda_{\Sigma},\qquad \sigma=\frac{1}{4}\ln\left[\frac{3m}{g}\right],\nonumber \\ 
& &V_0=-\frac{20g^2}{3}\sqrt{\frac{m}{3g}},\qquad L^2=-\frac{5}{V_0}=\frac{3}{4g^2}\sqrt{\frac{3g}{m}}\, .
\end{eqnarray}
$V_0$ and $L$ denote the cosmological constant and the $AdS_6$ radius, respectively. By shifting the value of the dilaton to $\sigma=0$ at the vacuum, we can set $g=3m$ resulting in 
\begin{equation}
V_0=-20m^2\qquad \textrm{and}\qquad L=\frac{1}{2m}
\end{equation}
in which we have chosen $m>0$.
\section{Supersymmetric $AdS_6$ black holes with $\Sigma\times \widetilde{\Sigma}$ horizons}\label{AdS2_Sigma_Sigma}
In this section, we consider black hole solutions with the near horizon geometry of the form $AdS_2\times \Sigma\times \widetilde{\Sigma}$ for $\Sigma$ and $\widetilde{\Sigma}$ being Riemann surfaces. The metric ansatz is given by
\begin{equation}
ds^2=-e^{2f(r)}dt^2+dr^2+e^{2h(r)}(d\theta^2+F_\kappa(\theta)^2 d\phi^2)+e^{2\tilde{h}(r)}(d\tilde{\theta}^2+\tilde{F}_{\tilde{\kappa}}(\tilde{\theta})^2 d\tilde{\phi}^2)
\end{equation}
with 
\begin{equation}
F_\kappa(\theta)=\begin{cases}
  \sin\theta,  & \kappa=1\phantom{-}\quad \textrm{for}\quad \Sigma^2=S^2 \\
  \theta,  & \kappa=0\phantom{-}\quad \textrm{for}\quad \Sigma^2=T^2\\
  \sinh\theta,  & \kappa=-1\quad \textrm{for}\quad \Sigma^2=H^2
\end{cases}\label{F_def}
\end{equation}
and similarly for $\tilde{F}_{\tilde{\kappa}}(\tilde{\theta})$. 
\\
\indent With the vielbein of the form 
\begin{eqnarray}
& &e^{\hat{t}}=e^fdt,\qquad e^{\hat{r}}=dr,\qquad e^{\hat{\theta}}=e^hd\theta,\nonumber \\
& &e^{\hat{\phi}}=e^hF_\kappa(\theta) d\phi,\qquad e^{\hat{\tilde{\theta}}}=e^{\tilde{h}}d\tilde{\theta},\qquad e^{\hat{\tilde{\phi}}}=e^{\tilde{h}}\tilde{F}_{\tilde{\kappa}}(\tilde{\theta})d\tilde{\phi},
\end{eqnarray}
we find the following non-vanishing components of the spin connection
\begin{eqnarray}
& &{\omega^{\hat{t}}}_{\hat{r}} =f'e^{\hat{t}},\qquad
{\omega^{\hat{\theta}}}_{\hat{r}}=h'e^{\hat{\theta}},\qquad {\omega^{\hat{\phi}}}_{\hat{r}}=h'e^{\hat{\phi}},\qquad {\omega^{\hat{\tilde{\theta}}}}_{\hat{r}}=\tilde{h}'e^{\hat{\tilde{\theta}}}, \nonumber \\
& &{\omega^{\hat{\tilde{\phi}}}}_{\hat{r}}=\tilde{h}'e^{\hat{\tilde{\phi}}},\qquad {\omega^{\hat{\phi}}}_{\hat{\theta}}=\frac{F'_\kappa(\theta)}{F_\kappa(\theta)}e^{-h}e^{\hat{\phi}},\qquad
{\omega^{\hat{\tilde{\phi}}}}_{\hat{\tilde{\theta}}}=\frac{\tilde{F}'_{\tilde{\kappa}}(\tilde{\theta})}{\tilde{F}_{\tilde{\kappa}}(\tilde{\theta})}e^{-\tilde{h}}e^{\hat{\tilde{\phi}}}\, .
\end{eqnarray}
We will use $'$ to denote $r$-derivatives except for $F'_\kappa(\theta)=\frac{dF_\kappa(\theta)}{d\theta}$ and $\tilde{F}'_{\tilde{\kappa}}(\tilde{\theta})=\frac{d\tilde{F}_{\tilde{\kappa}}(\tilde{\theta})}{d\tilde{\theta}}$. 
\\
\indent We will consider solutions with $SO(2)\times U(1)$ symmetry with $SO(2)\subset SO(3)\subset ISO(3)$. As shown in \cite{ISO3_flow}, there are six singlet scalars from $SO(4,4)/SO(4)\times SO(4)$ coset. These scalars can be described by the coset representative
\begin{equation}
L=e^{\phi_0Y_{03}}e^{\phi_1(Y_{11}+Y_{22})}e^{\phi_2Y_{33}}e^{\phi_3(Y_{12}-Y_{21})}e^{\phi_4Y_{04}}e^{\phi_5Y_{34}}\, .\label{L_SO2}
\end{equation}
In this equation, we have used the $SO(4,4)$ non-compact generators of the form
\begin{eqnarray}
Y_{\alpha I}=e^{\alpha,I+3}+e^{I+3,\alpha}
\end{eqnarray}  
for 
\begin{equation}
(e^{\Lambda \Sigma})_{\Gamma \Pi}=\delta^\Lambda_{
\Gamma}\delta^\Sigma_{\Pi},\qquad \Lambda, \Sigma,\Gamma,
\Pi=0,\ldots ,7\, .
\end{equation}
\indent We now perform a topological twist by turning on the following $SO(2)\times U(1)$ gauge fields
\begin{equation}
A^3=aF'_\kappa(\theta)d\phi+\tilde{a}\tilde{F}'_{\tilde{\kappa}}(\tilde{\theta})d\tilde{\phi}\qquad \textrm{and}\qquad A^7=bF'_\kappa(\theta)d\phi+\tilde{b}\tilde{F}'_{\tilde{\kappa}}(\tilde{\theta})d\tilde{\phi}\, .\label{SO2_U1_gauge_field}
\end{equation}
The composite connection is given by, see also \cite{ISO3_defect}, 
\begin{equation}
Q_{AB}=-\frac{i}{2}gA^3 \sigma^3_{AB}\, .\label{SO2_U1_composite}
\end{equation}
It should be noted that the field equations for the gauge fields given by the ansatz \eqref{SO2_U1_gauge_field} implies that $\phi_3=0$. Following \cite{AdS6_BH}, we will implement the relevant topological twist by imposing the following projectors and the twist conditions
\begin{eqnarray}
& &\gamma_{\hat{\theta}\hat{\phi}}\epsilon_A=\mp i\sigma^3_{AB}\epsilon^B,\qquad \gamma_{\hat{\tilde{\theta}}\hat{\tilde{\phi}}}\epsilon_A=\mp i\sigma^3_{AB}\epsilon^B,\nonumber \\
& &ga=\pm 1\qquad \textrm{and}\qquad g\tilde{a}=\pm 1\, .\label{SO2_twist}
\end{eqnarray}
It is also useful to note the gauge field strength tensors
\begin{eqnarray}
& &F^3=-\kappa aF_\kappa(\theta)d\theta\wedge d\phi-\tilde{\kappa}\tilde{a}\tilde{F}_{\tilde{\kappa}}(\tilde{\theta})d\tilde{\theta}\wedge d\tilde{\phi},\nonumber \\
& &F^7=-\kappa bF_\kappa(\theta)d\theta\wedge d\phi-\tilde{\kappa}\tilde{b}\tilde{F}_{\tilde{\kappa}}(\tilde{\theta})d\tilde{\theta}\wedge d\tilde{\phi}\, .
\end{eqnarray}
\indent Unlike the $AdS_4\times \Sigma$ and $AdS_3\times \mc{M}_3$ solutions considered in \cite{ISO3_defect}, in the present case, we cannot consistently set the two-form field to zero. Using the two-form field equation given in \cite{AdS6_BH}, we find that $H_{\mu\nu\rho}=0$ with the only non-vanishing component of the two-form field given by
\begin{equation}
B_{\hat{t}\hat{r}}=\frac{1}{8}\frac{\kappa\tilde{\kappa}(a\tilde{a}-b\tilde{b})}{m^2\mc{N}_{00}}e^{2\sigma-2h-2\tilde{h}}\, .
\end{equation}
We refer to \cite{AdS6_BH} for more detail on this issue. Finally, we also need to impose a projector involving $\gamma^{\hat{r}}$ of the form 
\begin{equation}
\gamma^{\hat{r}}\epsilon_A=-\epsilon_A\label{gamma_r_proj}
\end{equation}
with the sign choice chosen such that the $AdS_6$ vacuum appears at $r\rightarrow \infty$. As in \cite{AdS6_BH_Zaffaroni,AdS6_BH_Minwoo,AdS6_BH}, consistency of the BPS equations requires $\phi_0=0$. Moreover, in the present case, we also need $\phi_4=0$ in order to obtain a consistent set of BPS equations. 
\\
\indent Taking all of these into account and choosing the upper sign choice in \eqref{SO2_twist} for definiteness, we find the following BPS equations
\begin{eqnarray}
\phi_1'&=&2ge^{\sigma+2\phi_1}\cosh\phi_5\sinh\phi_2,\label{eq11}\\
\phi_2'&=&2ge^\sigma\textrm{sech}\phi_5\left[(e^{2\phi_1}-1)\cosh\phi_2-\sinh\phi_2\right]\nonumber \\
& &-\frac{1}{2}e^{-\sigma-2h-2\tilde{h}}\textrm{sech}\phi_5\sinh\phi_2(a\kappa e^{2\tilde{h}}+\tilde{a}\tilde{\kappa}e^{2h}),\label{eq12}\\
\phi_5'&=&-2ge^\sigma \sinh\phi_5\left[\cosh\phi_2-\sinh\phi_2(e^{2\phi_1}-1)\right]+\frac{1}{2}e^{-\sigma-2h-2\tilde{h}}\left[\kappa e^{2\tilde{h}}(b\cosh\phi_5\right.\nonumber \\
& &\left. -a\cosh\phi_2\sinh\phi_5) +e^{2h}\tilde{\kappa}(\tilde{b}\cosh\phi_5-\tilde{a}\cosh\phi_2\sinh\phi_5)\right],\label{eq13}
\end{eqnarray}
\begin{eqnarray}
\sigma'&=&-\frac{1}{2}ge^\sigma\cosh\phi_5\left[\cosh\phi_2-\sinh\phi_2(e^{2\phi_1}-1)\right]+\frac{1}{32}\frac{(a\tilde{a}-b\tilde{b})\kappa\tilde{\kappa}}{m}e^{\sigma-2h-2\tilde{h}}\nonumber \\
& &+\frac{3}{2}me^{-3\sigma}+\frac{1}{8}e^{-\sigma-2h-2\tilde{h}}\left[e^{2\tilde{h}}\kappa (a\cosh\phi_2\cosh\phi_5-b\sinh\phi_5)\right. \nonumber \\
& &\left.+e^{2h}\tilde{\kappa}(\tilde{a}\cosh\phi_2\cosh\phi_5-\tilde{b}\sinh\phi_5)\right],\\
h'&=&\frac{1}{2}ge^\sigma\cosh\phi_5\left[\cosh\phi_2-\sinh\phi_2(e^{2\phi_1}-1)\right]+\frac{1}{32}\frac{(b\tilde{b}-a\tilde{a})\kappa\tilde{\kappa}}{m}e^{\sigma-2h-2\tilde{h}}\nonumber \\
& &+\frac{1}{2}me^{-3\sigma}+\frac{1}{8}e^{-\sigma-2h-2\tilde{h}}\left[3\kappa e^{2\tilde{h}}(a\cosh\phi_2\cosh\phi_5-b\sinh\phi_5)\right. \nonumber \\
& &\left.-\tilde{\kappa}e^{2h}(\tilde{a}\cosh\phi_2\cosh\phi_5-\tilde{b}\sinh\phi_5)\right],\\
\tilde{h}'&=&\frac{1}{2}ge^\sigma\cosh\phi_5\left[\cosh\phi_2-\sinh\phi_2(e^{2\phi_1}-1)\right]+\frac{1}{32}\frac{(b\tilde{b}-a\tilde{a})\kappa\tilde{\kappa}}{m}e^{\sigma-2h-2\tilde{h}}\nonumber \\
& &+\frac{1}{2}me^{-3\sigma}+\frac{1}{8}e^{-\sigma-2h-2\tilde{h}}\left[-\kappa e^{2\tilde{h}}(a\cosh\phi_2\cosh\phi_5-b\sinh\phi_5)\right. \nonumber \\
& &\left.+3\tilde{\kappa}e^{2h}(\tilde{a}\cosh\phi_2\cosh\phi_5-\tilde{b}\sinh\phi_5)\right],\\
f'&=&\frac{1}{2}ge^\sigma\cosh\phi_5\left[\cosh\phi_2-\sinh\phi_2(e^{2\phi_1}-1)\right]+\frac{3}{32}\frac{(b\tilde{b}-a\tilde{a})\kappa\tilde{\kappa}}{m}e^{\sigma-2h-2\tilde{h}}\nonumber \\
& &+\frac{1}{2}me^{-3\sigma}-\frac{1}{8}e^{-\sigma-2h-2\tilde{h}}\left[\kappa e^{2\tilde{h}}(a\cosh\phi_2\cosh\phi_5-b\sinh\phi_5)\right. \nonumber \\
& &\left.+\tilde{\kappa}e^{2h}(\tilde{a}\cosh\phi_2\cosh\phi_5-\tilde{b}\sinh\phi_5)\right].
\end{eqnarray}
We have also verified that these equations are compatible with the field equations.

\subsection{$AdS_2\times \Sigma\times \widetilde{\Sigma}$ vacua}
We now look for possible $AdS_2\times \Sigma\times \widetilde{\Sigma}$ vacua from the resulting BPS equations. At the horizon, we have the following conditions
\begin{equation}
\sigma'=h'=\tilde{h}'=\phi_i'=0,\qquad i=1,2,5,\qquad\textrm{and}\qquad  f'=\frac{1}{\ell} 
\end{equation}
with an $AdS_2$ radius $\ell$. Since the $\gamma^{\hat{r}}$ projector is not needed for constant scalars, the two projectors involving $\gamma_{\hat{\theta}\hat{\phi}}$ and $\gamma_{\hat{\tilde{\theta}}\hat{\tilde{\phi}}}$ in \eqref{SO2_twist} imply that the $AdS_2\times \Sigma\times \widetilde{\Sigma}$ solutions preserve four supercharges. From equation \eqref{eq11}, we always have $\phi_2=0$ at the fixed points. On the other hand, setting $\phi_2=0$ in equation \eqref{eq12} leads to $\phi_1=0$ to satisfy the $\phi_2'=0$ condition. Therefore, the existence of $AdS_2\times \Sigma\times \widetilde{\Sigma}$ solutions requires $\phi_1=\phi_2=0$. By the scalar masses given in \cite{ISO3_flow}, we also know that $\phi_1+\phi_2$ and $\phi_1-2\phi_2$ are dual to operators of dimensions $8$ and $6$, respectively. On the other hand, $\sigma$ and $\phi_5$ are both dual to operators of dimension $3$. We then see that the existence of $AdS_2$ fixed points requires scalars dual to irrelevant operators to vanish. 
\\
\indent From the BPS equations, we find two $AdS_2\times \Sigma\times \widetilde{\Sigma}$ solutions. These are given by
\begin{itemize}
\item $AdS_2$ critical point I:
\begin{eqnarray}
\phi_1&=&\phi_2=0,\qquad \phi_5=0,\qquad b=\tilde{b}=0, \nonumber\\
h&=&\frac{1}{2}\ln\left[-\frac{a\kappa}{4m}\sqrt{\frac{2m}{g}}\right],\qquad \tilde{h}=\frac{1}{2}\ln\left[-\frac{\tilde{a}\tilde{\kappa}}{4m}\sqrt{\frac{2m}{g}}\right],\nonumber \\
\sigma&=&\frac{1}{4}\ln\left[\frac{2m}{g}\right],\qquad
\ell=\frac{1}{2^{\frac{5}{4}}g^{\frac{3}{4}}m^{\frac{1}{4}}}\, .\label{AdS2_Sigma_Sigma1}
\end{eqnarray}  
\end{itemize}
We note that this solution only exists for $\kappa=-1$ and $\tilde{\kappa}=-1$, so this leads to $AdS_2\times H^2\times H^2$ geometry. It should also be pointed out that for $\phi_5=0$, consistency of equation \eqref{eq13} implies $b=\tilde{b}=0$.
\\
\indent Another $AdS_2\times \Sigma\times \widetilde{\Sigma}$ solution is given by
\begin{itemize}
\item $AdS_2$ critical point II:
\begin{eqnarray}
\phi_1&=&\phi_2=0,\qquad \phi_5=\varphi_0,\nonumber\\
 h&=&\frac{1}{2}\ln\left[\frac{\kappa e^{2\sigma}(b\sinh\varphi_0-a\cosh\varphi_0)}{4m}\right],\nonumber \\
 \tilde{h}&=&\frac{1}{2}\ln\left[\frac{\tilde{\kappa}e^{2\sigma}(\tilde{b}\sinh\varphi_0-\tilde{a}\cosh\varphi_0)}{4m}\right],\nonumber \\
\sigma&=&\frac{1}{4}\ln\left[\frac{m[3(b\tilde{b}-a\tilde{a})-(a\tilde{a}+b\tilde{b})\cosh2\varphi_0+(b\tilde{a}+a\tilde{b})\sinh2\varphi_0]}{2g\cosh^3\varphi_0(a-b\tanh\varphi_0)(\tilde{b}\tanh\varphi_0-\tilde{a})}\right],
\nonumber \\
\ell&=&\textrm{sech}\varphi_0\left[\frac{(a\cosh\varphi_0-b\sinh\varphi_0)(a\cosh2\varphi_0-b\sinh2\varphi_0)}{8g^3m[a^2-b^2+3(a^2+b^2)\cosh2\varphi_0-6ab\sinh2\varphi_0]}\right]^{\frac{1}{4}}\, \, \,
 \label{AdS2_Sigma_Sigma2}
\end{eqnarray}
with $\varphi_0$ given by a solution of the following equation
\begin{equation}
\tilde{b}=\frac{\tilde{a}\left[a(e^{2\varphi_0}-1)^3-b(e^{2\varphi_0}+1)^3\right]}{a(e^{2\varphi_0}+1)^3-b(e^{6\varphi_0}+9e^{4\varphi_0}-9e^{2\varphi_0}-1)}\, .
\end{equation} 
\end{itemize}
An explicit form for $\varphi_0$ can also be given. However, this is highly complicated, so we refrain from presenting it here. Unlike $AdS_2$ critical point I, in this case, there are more possibilities with $\kappa=\tilde{\kappa}=-1$, $\kappa=-1=-\tilde{\kappa}$ and $\kappa=1=-\tilde{\kappa}$. Therefore, there are $AdS_2\times H^2\times H^2$, $AdS_2\times S^2\times H^2$ and $AdS_2\times H^2\times S^2$ solutions depending on the values of $b$ and $\tilde{b}$. On the other hand, the values of $a$ and $\tilde{a}$ are fixed by the twist conditions \eqref{SO2_twist}. For $\kappa=\tilde{\kappa}=1$, the BPS equations do not admit $AdS_2$ fixed points for any values of $b$ and $\tilde{b}$. 

\subsection{Numerical black hole solutions}
\begin{figure}
         \centering
               \begin{subfigure}[b]{0.4\textwidth}
                 \includegraphics[width=\textwidth]{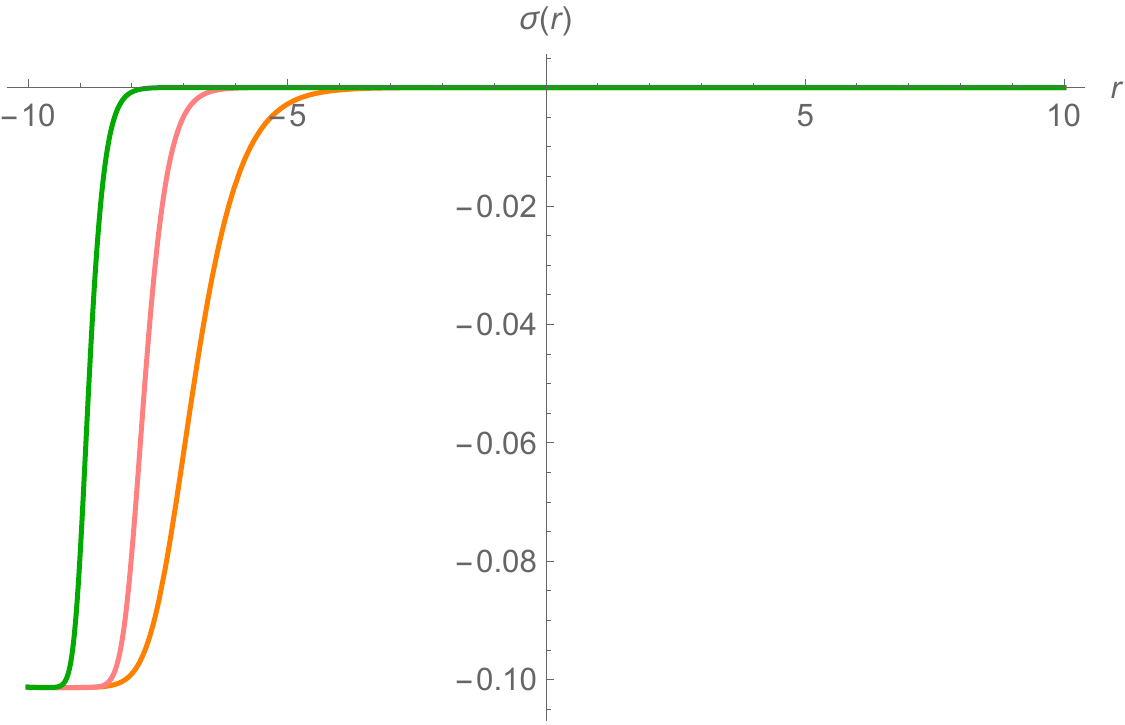}
                 \caption{Solutions for $\sigma(r)$}
         \end{subfigure}
          \begin{subfigure}[b]{0.4\textwidth}
                 \includegraphics[width=\textwidth]{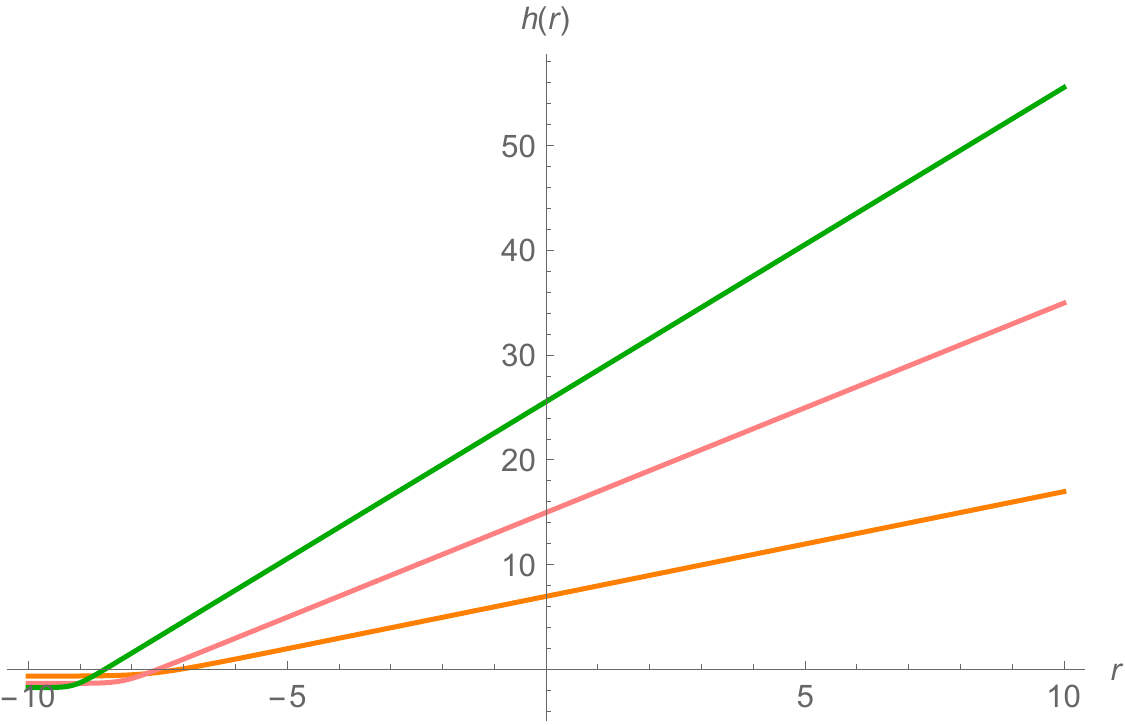}
                 \caption{Solutions for $h(r)$}
         \end{subfigure}\\
          \begin{subfigure}[b]{0.4\textwidth}
                 \includegraphics[width=\textwidth]{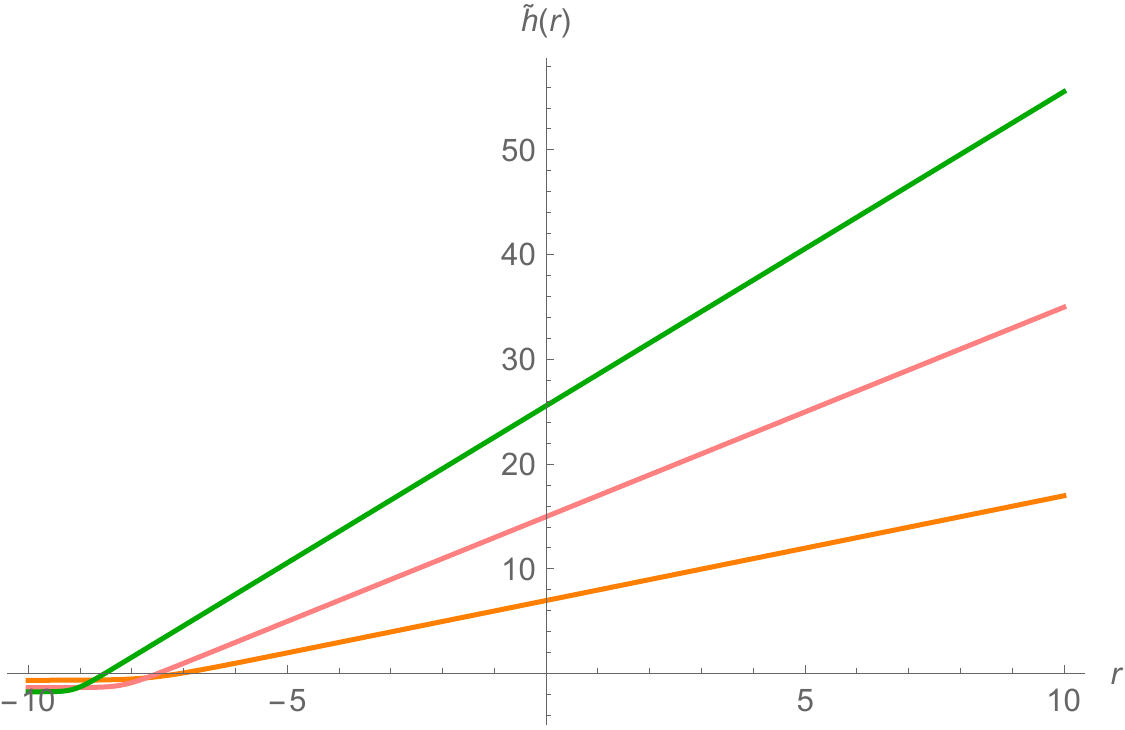}
                 \caption{Solutions for $\tilde{h}(r)$}
         \end{subfigure}
          \begin{subfigure}[b]{0.4\textwidth}
                 \includegraphics[width=\textwidth]{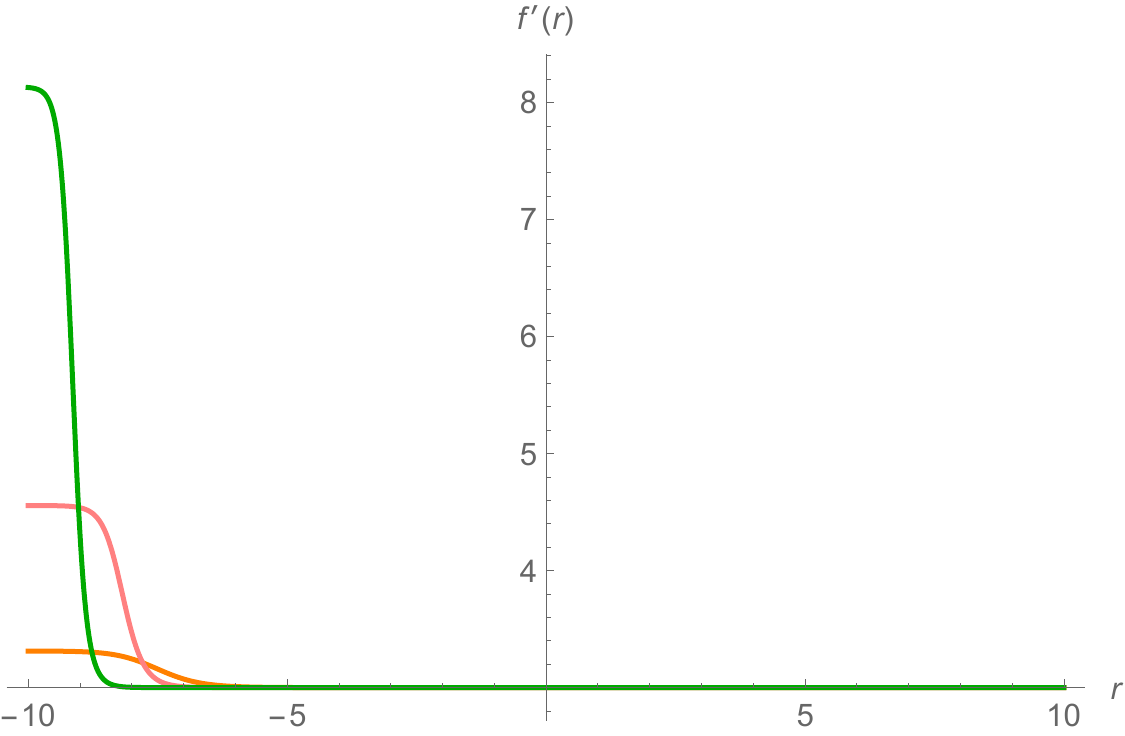}
                 \caption{Solutions for $f'(r)$}
         \end{subfigure}
\caption{Supersymmetric $AdS_6$ black holes interpolating between the $AdS_6$ vacuum and near horizon geometries $AdS_2\times H^2\times H^2$ (critical point I) for $m=\frac{1}{2}$ (orange), $m=1$ (pink), $m=\frac{3}{2}$ (green).}\label{fig1}
 \end{figure}

We now numerically find black hole solutions interpolating between an asymptotically locally $AdS_6$ vacuum and near horizon $AdS_2\times \Sigma\times \widetilde{\Sigma}$ geometries identified previously. We first consider solutions flowing to $AdS_2$ critical point I which is given by $AdS_2\times H^2\times H^2$ geometry. We will choose the following values of the gauge coupling constant $g=3m$ and set $\phi_1=\phi_2=\phi_5=0$, $\kappa=\tilde{\kappa}=-1$ and $b=\tilde{b}=0$. These solutions do not involve any fields from the fourth vector multiplet and can be regarded as solutions of $F(4)$ gauged supergravity coupled to three vector multiplets with $ISO(3)$ gauge group. This is a consistent truncation of the $ISO(3)\times U(1)$ $F(4)$ gauged supergravity and, as pointed out in \cite{Henning_Malek_AdS7_6}, it is possible to obtain this gauged supergravity from a consistent truncation of type IIB theory on $S^2\times \Sigma$ as well. Examples of numerical solutions with different values of $m$ are shown in figure \ref{fig1}. 
\\
\indent The corresponding black hole entropy is given by the relation
\begin{equation}
S_{\textrm{BH}}=\frac{1}{4G_{\textrm{N}}}e^{2h+2\tilde{h}}\textrm{vol}(\Sigma)\textrm{vol}(\widetilde{\Sigma})
\end{equation}
with the volume of a genus-$\mathfrak{g}$ Riemann surface given by
\begin{equation}
\textrm{vol}(\Sigma_{\mathfrak{g}})=2\pi \eta_{\mathfrak{g}}\qquad \textrm{and}\qquad \eta_{\mathfrak{g}}=\begin{cases}
  2|\mathfrak{g}-1|,  &\quad  \textrm{for}\quad \mathfrak{g}\neq 1 \\
  1,  & \quad \textrm{for}\quad \mathfrak{g}=1
\end{cases}\, .
\end{equation}
For the present case of black holes with $H^2\times H^2$ horizon, we find 
\begin{equation}
S_{\textrm{BH}}=\frac{\pi^2a\tilde{a}\eta_{\mathfrak{g}}\eta_{\tilde{\mathfrak{g}}}}{24m^2G_{\textrm{N}}}\, .
\end{equation}    

\begin{figure}
         \centering
               \begin{subfigure}[b]{0.45\textwidth}
                 \includegraphics[width=\textwidth]{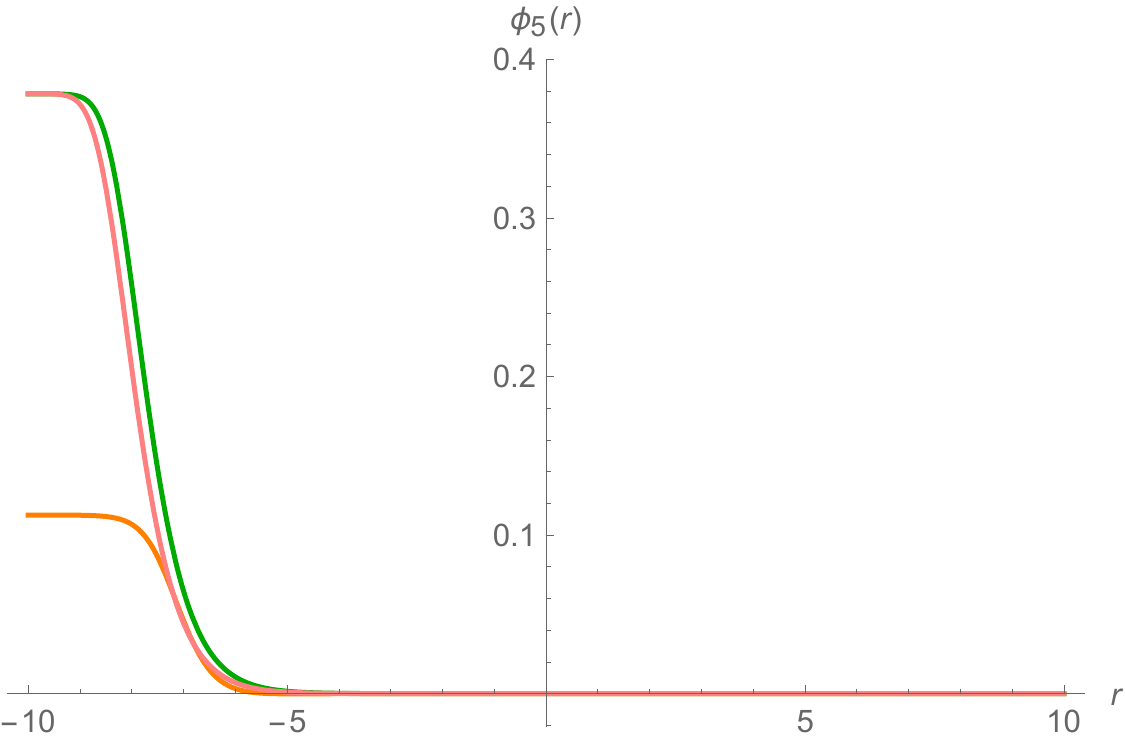}
                 \caption{Solutions for $\phi_5(r)$}
         \end{subfigure}
          \begin{subfigure}[b]{0.45\textwidth}
                 \includegraphics[width=\textwidth]{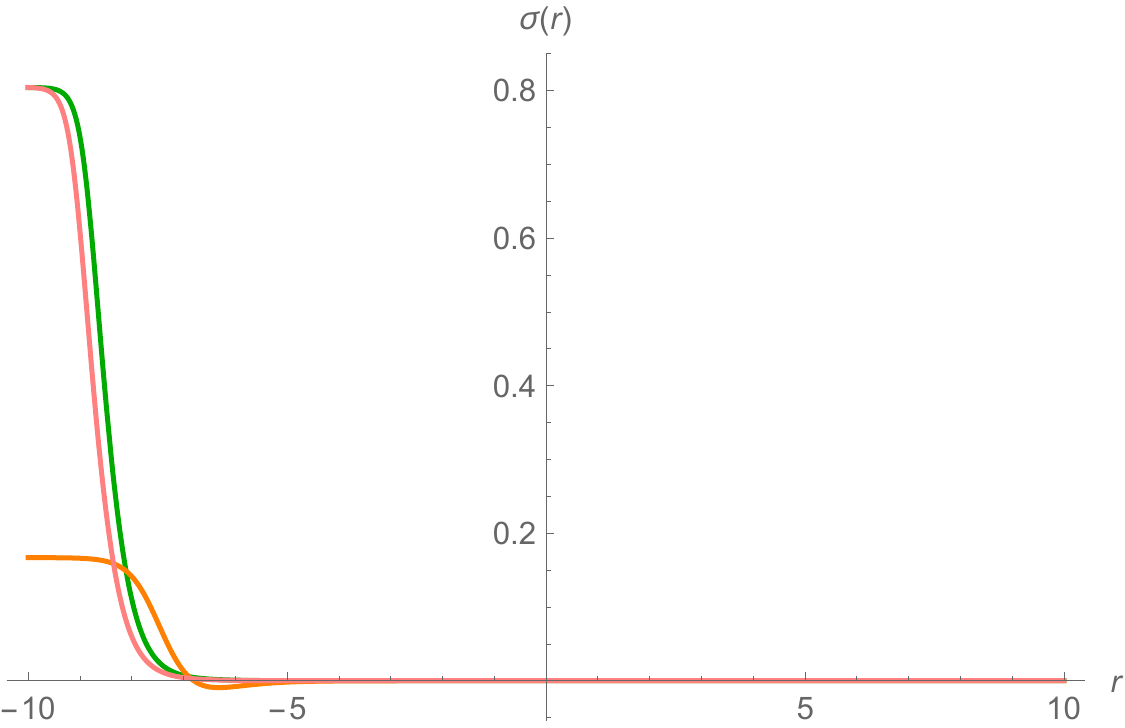}
                 \caption{Solutions for $\sigma(r)$}
         \end{subfigure}\\
          \begin{subfigure}[b]{0.45\textwidth}
                 \includegraphics[width=\textwidth]{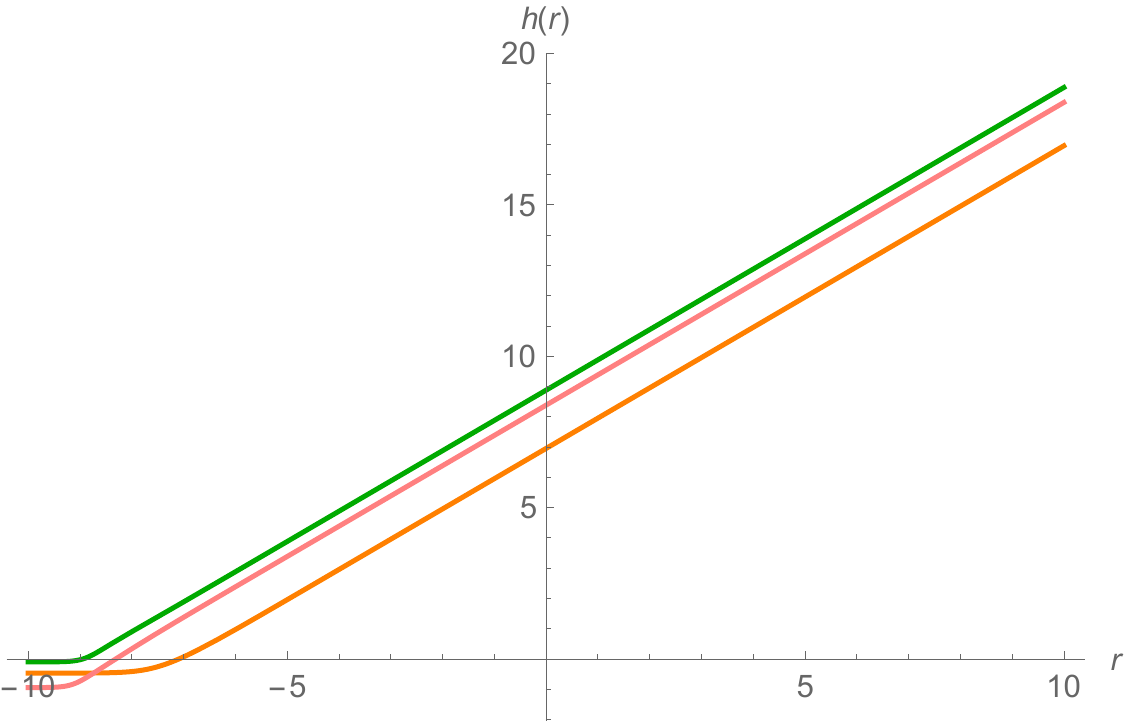}
                 \caption{Solutions for $h(r)$}
         \end{subfigure}
                   \begin{subfigure}[b]{0.45\textwidth}
                 \includegraphics[width=\textwidth]{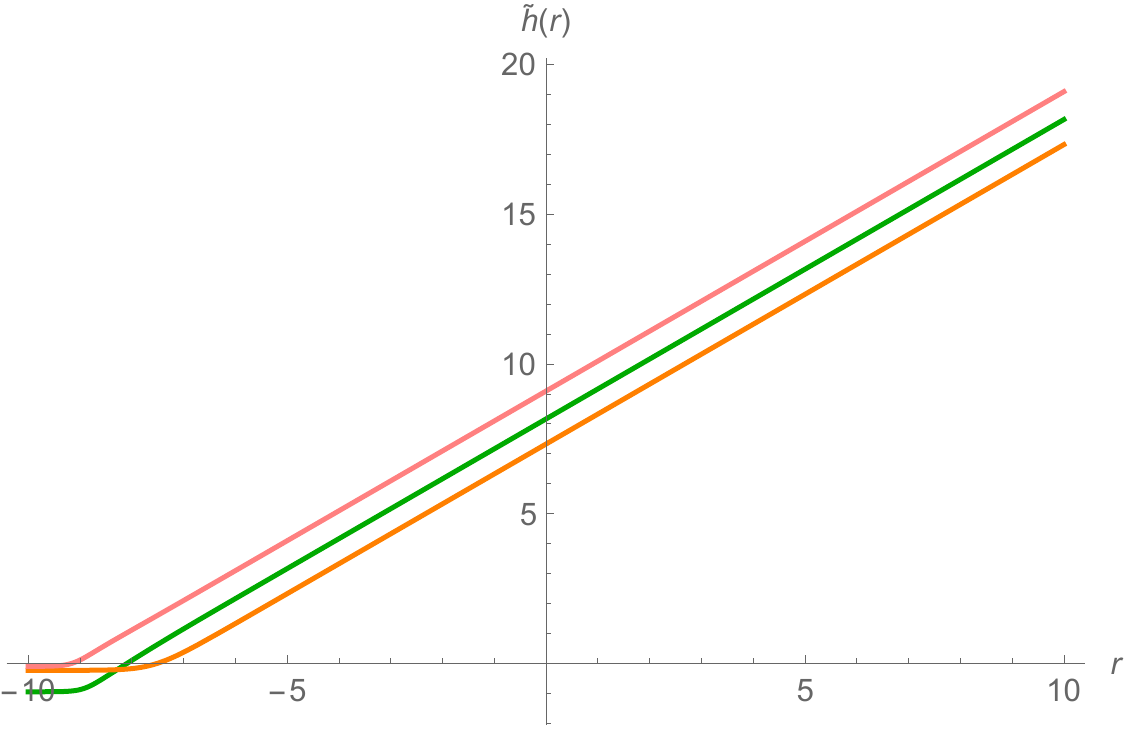}
                 \caption{Solutions for $\tilde{h}(r)$}
           \end{subfigure} \\
          \begin{subfigure}[b]{0.45\textwidth}
                 \includegraphics[width=\textwidth]{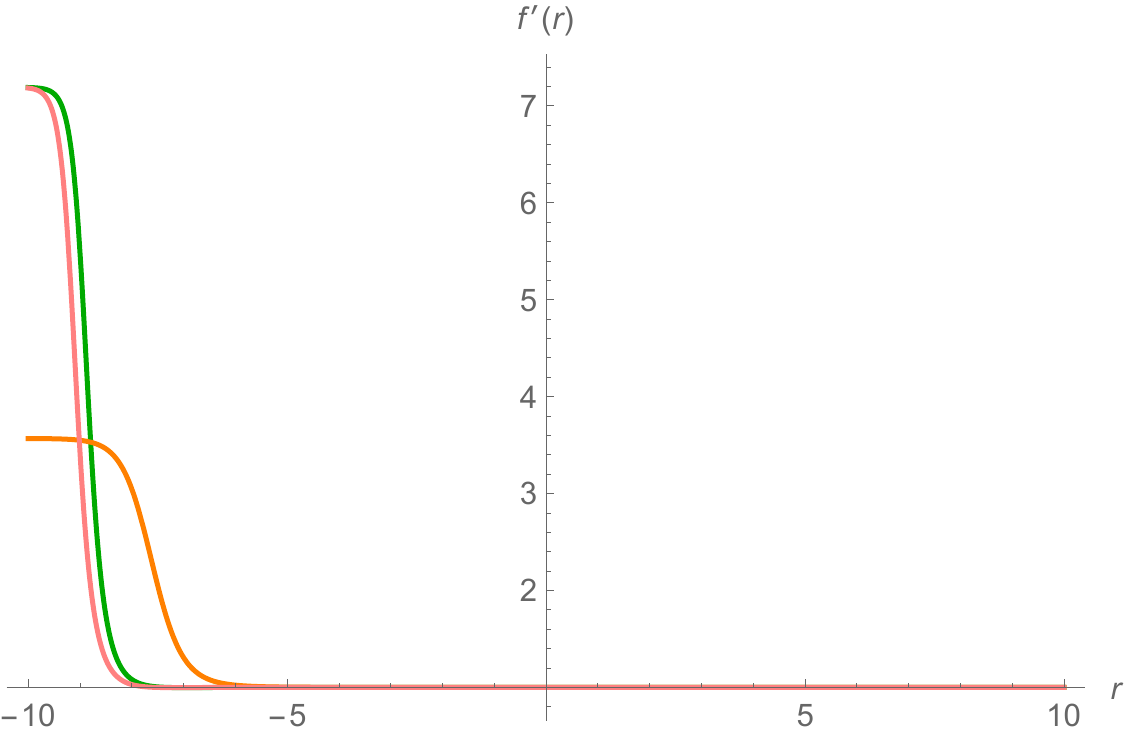}
                 \caption{Solutions for $f'(r)$}
         \end{subfigure}
\caption{Supersymmetric $AdS_6$ black holes with $AdS_2\times H^2\times H^2$ (orange), $AdS_2\times S^2\times H^2$ (pink) and $AdS_2\times H^2\times S^2$ (green) near horizon geometries given by critical point II.}\label{fig2}
\end{figure}

\indent We now move to more complicated black hole solutions with near horizon geometries given by $AdS_2\times \Sigma\times \widetilde{\Sigma}$ critical point II. We will set $\phi_1=\phi_2=0$ and choose $g=3m$ and $m=\frac{1}{2}$ leading to an $AdS_6$ vacuum of unit radius. Examples of black hole solutions are shown in figure \ref{fig2}. In the figure, there is a solution that flows from the $AdS_6$ vacuum to $AdS_2\times H^2\times S^2$ geometry (green curve) and also a solution interpolating between the $AdS_6$ vacuum and $AdS_2\times S^2\times H^2$ geometry (pink curve). These solutions are obtained by choosing $(b,\tilde{b})=(\pm 1,\pm 2)$ and $(b,\tilde{b})=(\pm 2,\pm1)$, respectively. The solution interpolating between the $AdS_6$ vacuum and $AdS_2\times H^2\times H^2$ geometry is shown by the orange curve. This solution is obtained by setting $b=\pm1$ and $\tilde{b}=\mp 2$. Examples of $AdS_2\times H^2\times H^2$ geometries can also be obtained by choosing $b=\pm2$ and $\tilde{b}=\mp 1$.
\\
\indent The corresponding black hole entropy in this case is given by
 \begin{equation}
S_{\textrm{BH}}=\frac{\pi^2\kappa\tilde{\kappa}\eta_{\mathfrak{g}}\eta_{\tilde{\mathfrak{g}}}\left[(a\tilde{a}+b\tilde{b})\cosh\varphi_0
+(a\tilde{a}-2b\tilde{b})\textrm{sech}\varphi_0-(b\tilde{a}+a\tilde{b})\sinh\varphi_0\right]}{48m^2G_{\textrm{N}}}\, .
\end{equation}    

\section{Supersymmetric $AdS_6$ black holes with $\mc{M}_4$ horizons}\label{AdS2_M4}
In this section, we repeat the same analysis for $AdS_6$ black holes with near horizon geometries of the form $AdS_2\times \mc{M}_4$ for $\mc{M}_4$ being an Einstein four-manifold. We will consider two types of $\mc{M}_4$ namely a Kahler four-cycle and a Cayley four-cycle as in \cite{AdS6_BH}. The procedures are essentially the same as in \cite{AdS6_BH_Minwoo} and \cite{AdS6_BH}, so we will mainly give relevant formulae that lead to the presented results and refer to \cite{AdS6_BH} for more detail.  

\subsection{Black holes with Kahler four-cycle horizon}
\begin{figure}
         \centering
               \begin{subfigure}[b]{0.4\textwidth}
                 \includegraphics[width=\textwidth]{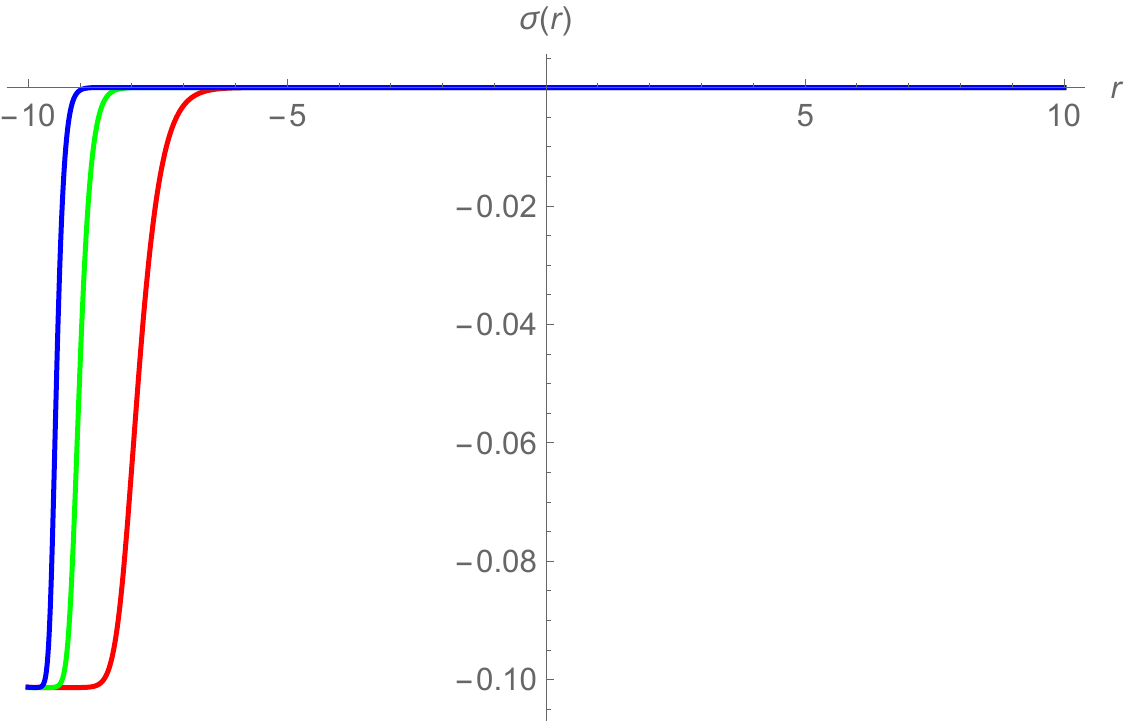}
                 \caption{Solutions for $\sigma(r)$}
         \end{subfigure}
          \begin{subfigure}[b]{0.4\textwidth}
                 \includegraphics[width=\textwidth]{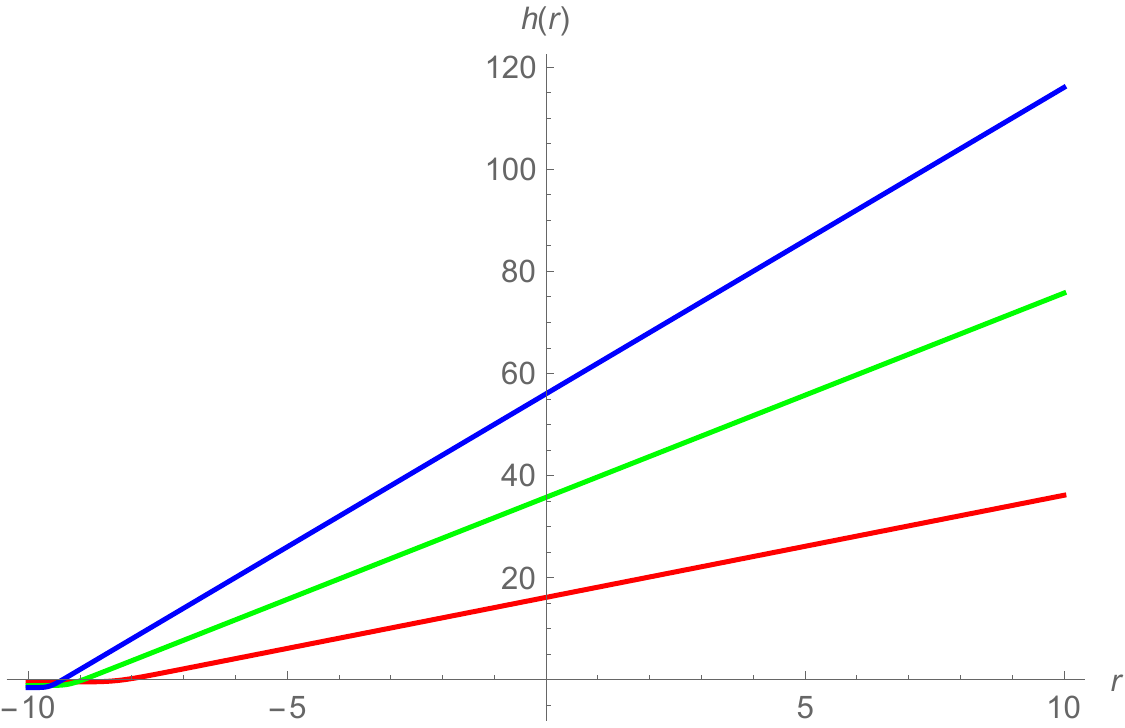}
                 \caption{Solutions for $h(r)$}
         \end{subfigure}\\
          \begin{subfigure}[b]{0.4\textwidth}
                 \includegraphics[width=\textwidth]{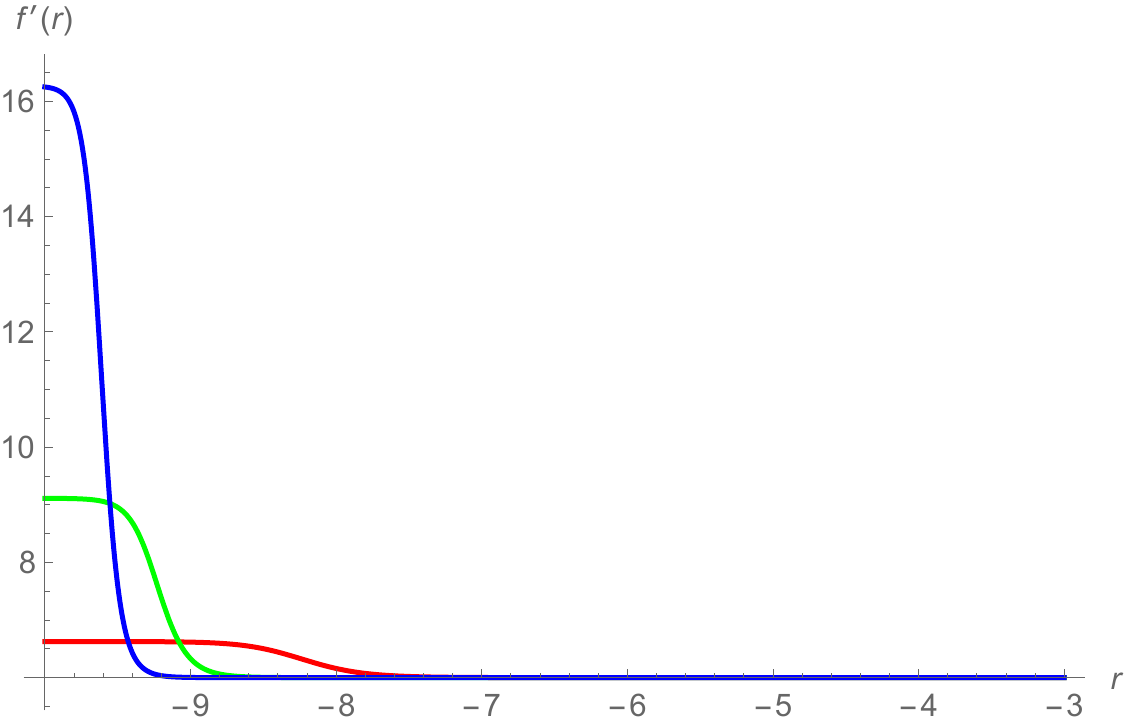}
                 \caption{Solutions for $f'(r)$}
         \end{subfigure}
\caption{Supersymmetric $AdS_6$ black holes interpolating between the $AdS_6$ vacuum and the near horizon geometry $AdS_2\times \mc{M}_{\textrm{K}4}^-$ (critical point III) for $m=1$ (red), $m=2$ (green), $m=3$ (blue).}\label{fig3}
 \end{figure}

\begin{figure}
         \centering
               \begin{subfigure}[b]{0.4\textwidth}
                 \includegraphics[width=\textwidth]{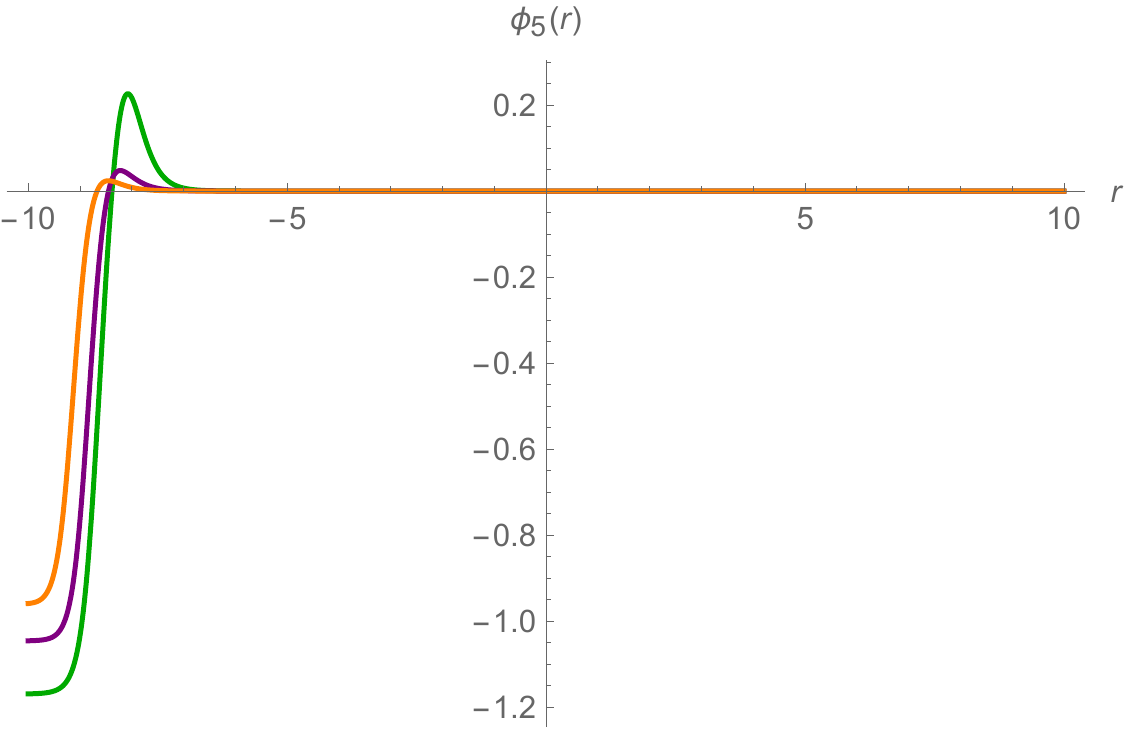}
                 \caption{Solutions for $\phi_5(r)$}
         \end{subfigure}
          \begin{subfigure}[b]{0.4\textwidth}
                 \includegraphics[width=\textwidth]{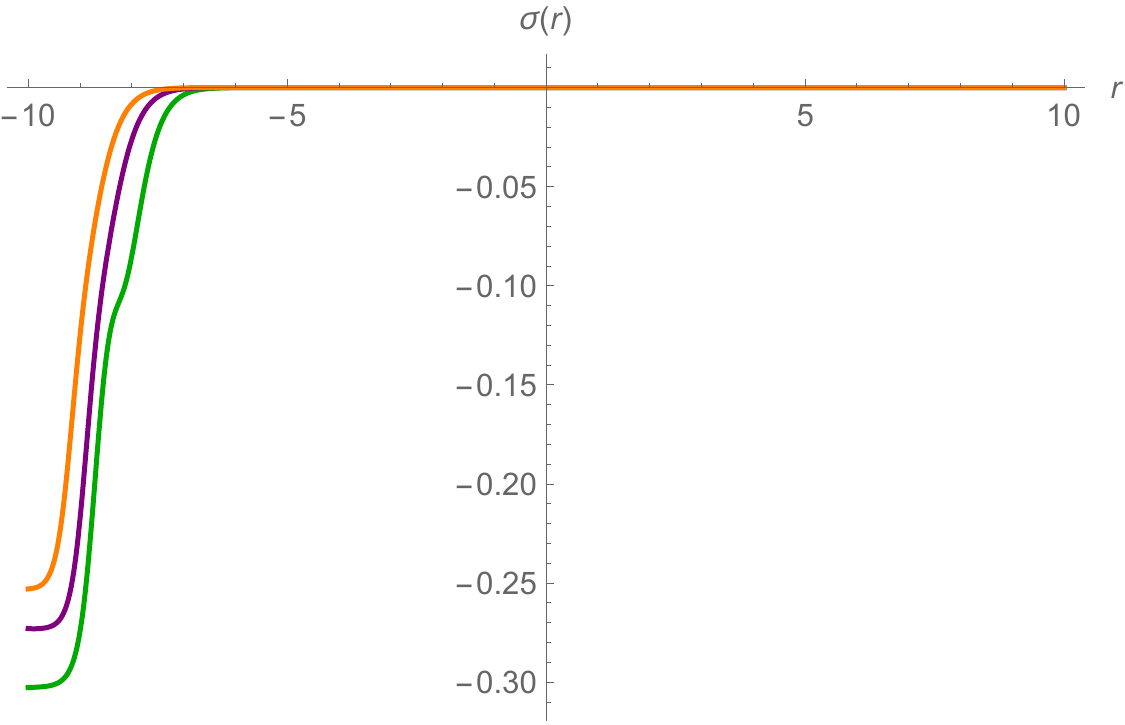}
                 \caption{Solutions for $\sigma(r)$}
         \end{subfigure}\\
          \begin{subfigure}[b]{0.4\textwidth}
                 \includegraphics[width=\textwidth]{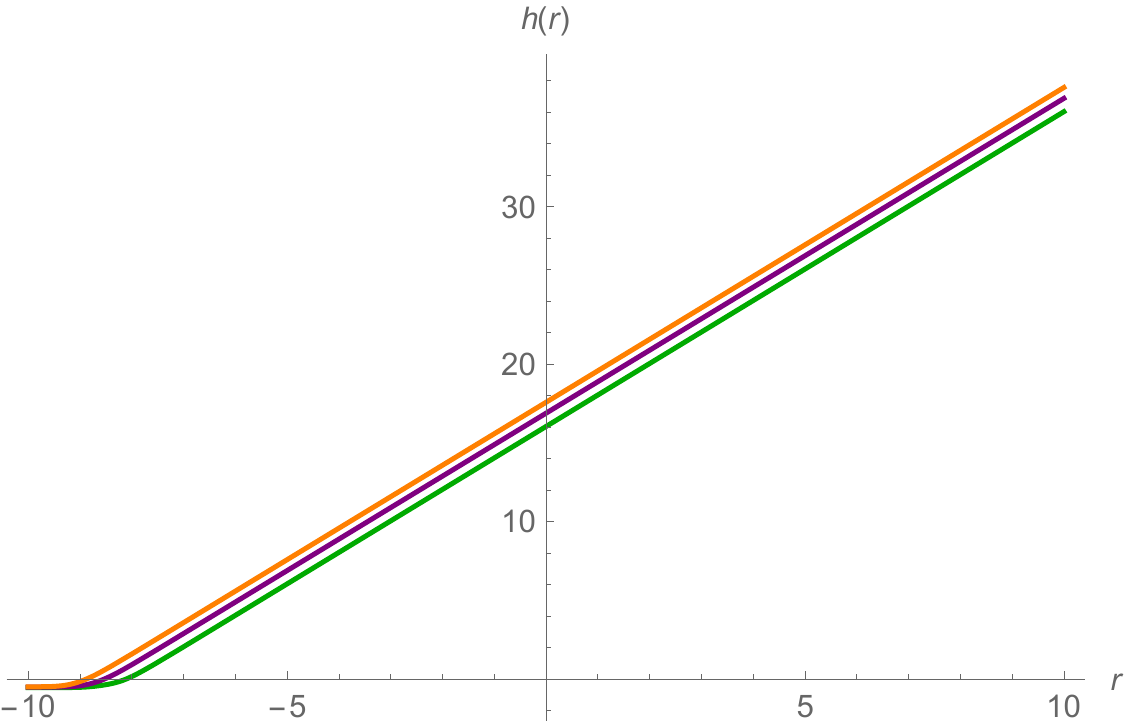}
                 \caption{Solutions for $h(r)$}
         \end{subfigure}
          \begin{subfigure}[b]{0.4\textwidth}
                 \includegraphics[width=\textwidth]{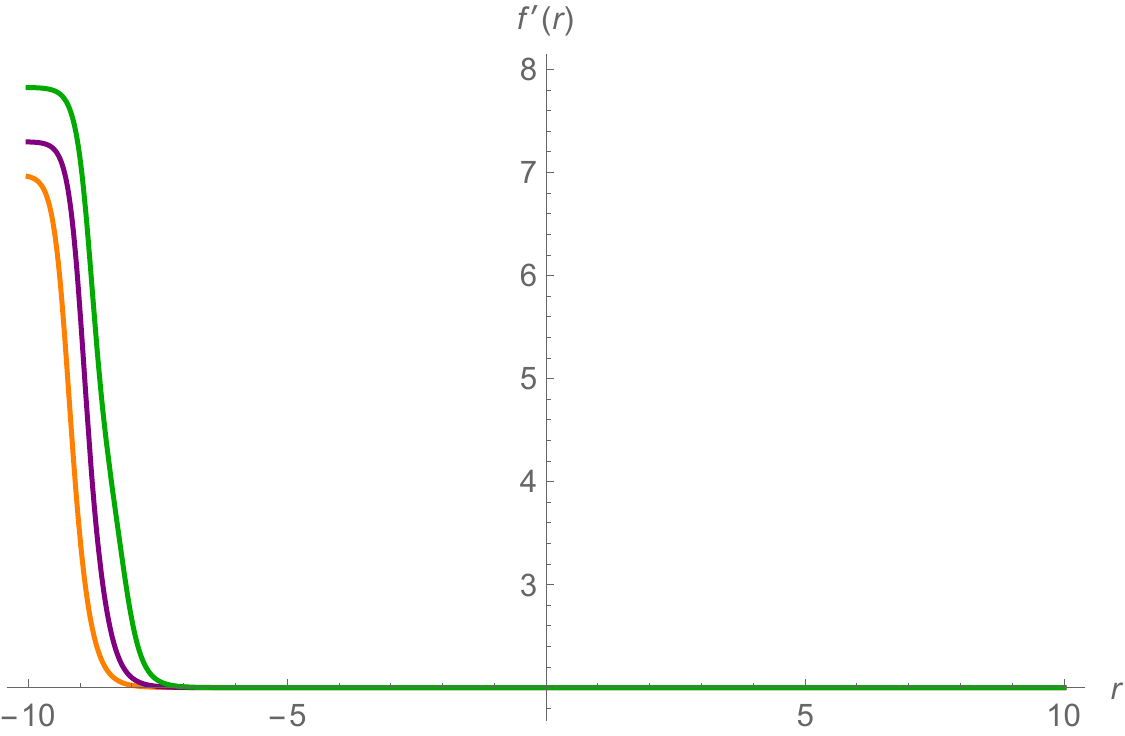}
                 \caption{Solutions for $f'(r)$}
         \end{subfigure}
\caption{Supersymmetric $AdS_6$ black holes interpolating between the $AdS_6$ vacuum and the near horizon geometry $AdS_2\times \mc{M}_{\textrm{K}4}^-$ (critical point IV) for $b=-\frac{1}{8}$ (green), $b=-\frac{1}{10}$ (purple), $b=-\frac{1}{12}$ (orange).}\label{fig4}
 \end{figure}

For $\mc{M}_4$ being a Kahler four-cycle with a $U(2)\sim U(1)\times SU(2)$ holonomy. We will perform the topological twist along the $U(1)$ part by turning on the $SO(2)\times U(1)$ gauge field as in the previous section. The $SO(4,4)/SO(4)\times SO(4)$ coset representative is again given by \eqref{L_SO2}, and the metric ansatz takes the form
\begin{equation}
ds^2=-e^{2f(r)}dt^2+dr^2+e^{2h(r)}ds^2_{\mc{M}_{\textrm{K}4}}
\end{equation}
with the metric on $\mc{M}_{\textrm{K}4}$ given by
\begin{equation}
ds^2_{\mc{M}_{\textrm{K}4}}=\frac{1}{f_\kappa^2(\rho)}\left[d\rho^2+\rho^2f_\kappa(\rho)(\tau_1^2+\tau^2_2)+\rho^2\tau_3^2\right].
\end{equation}
The function $f_\kappa(\rho)=1+\kappa \rho^2$, and $\tau_i$, $i=1,2,3$, are $SU(2)$ left-invariant one-forms with the normalization
\begin{equation}
d\tau_i=\epsilon_{ijk}\tau_j\wedge \tau_k\, .
\end{equation}
With the following choice of vielbein
\begin{eqnarray}
& &e^{\hat{t}}=e^fdt,\qquad e^{\hat{r}}=dr,\qquad e^{\hat{1}}=\frac{e^h\rho}{\sqrt{f_\kappa(\rho)}}\tau_1,\nonumber \\
& &e^{\hat{2}}=\frac{e^h\rho}{\sqrt{f_\kappa(\rho)}}\tau_2,\qquad e^{\hat{3}}=\frac{e^h\rho}{f_\kappa(\rho)}\tau_3,\qquad e^{\hat{4}}=\frac{e^h}{f_\kappa(\rho)}d\rho,
\end{eqnarray}
non-vanishing components of the spin connection are given by
\begin{eqnarray}
& &{\omega^{\hat{t}}}_{\hat{r}}=f'e^{\hat{t}},\qquad {\omega^{\hat{\alpha}}}_{\hat{r}}=h'e^{\hat{\alpha}},\quad \hat{\alpha}=1,2,3,4,\nonumber \\
& &{\omega^{\hat{1}}}_{\hat{4}}={\omega^{\hat{2}}}_{\hat{3}}=\frac{e^{-h}}{\rho}e^{\hat{1}},\qquad
{\omega^{\hat{2}}}_{\hat{4}}={\omega^{\hat{3}}}_{\hat{1}}=\frac{e^{-h}}{\rho}e^{\hat{2}},\nonumber \\ 
& &{\omega^{\hat{1}}}_{\hat{2}}=\frac{e^{-h}}{\rho}(2\kappa \rho^2+1)e^{\hat{3}},\qquad {\omega^{\hat{4}}}_{\hat{3}}=\frac{e^{-h}}{\rho}(\kappa \rho^2-1)e^{\hat{3}}\, .
\end{eqnarray}
To implement the topological twist, we turn on the $SO(2)\times U(1)$ gauge fields of the form
\begin{equation}
A^3=3a\kappa\rho e^{-h}e^{\hat{3}}\qquad \textrm{and}\qquad A^7=3b\kappa \rho e^{-h}e^{\hat{3}}\, .
\end{equation} 
The composite connection is still given by \eqref{SO2_U1_composite}. We then impose the projectors 
\begin{equation}
\gamma_{\hat{1}\hat{2}}\epsilon_A=-\gamma_{\hat{3}\hat{4}}\epsilon_A=i\sigma^3_{AB}\epsilon^B\label{Proj_Kahler4}
\end{equation}
and the twist condition
\begin{equation}
ga=1\, .
\end{equation}
In this case, the gauge field strength tensors are given by
\begin{equation}
F^3=6\kappa ae^{-2h}(e^{\hat{1}}\wedge e^{\hat{2}}-e^{\hat{3}}\wedge e^{\hat{4}})\qquad \textrm{and}\qquad F^7=6\kappa be^{-2h}(e^{\hat{1}}\wedge e^{\hat{2}}-e^{\hat{3}}\wedge e^{\hat{4}})
\end{equation}
which imply the non-vanishing component of the two-form field
\begin{equation}
B_{\hat{t}\hat{r}}=\frac{9}{2}\frac{\kappa^2e^{2\sigma-4h}}{m^2\mc{N}_{00}}(b^2-a^2).
\end{equation}
\indent As in the previous section, consistency requires $\phi_0=\phi_3=\phi_4=0$. Using the $\gamma^{\hat{r}}$ projector given in \eqref{gamma_r_proj}, we find the following BPS equations
\begin{eqnarray}
\phi_1'&=&2ge^{\sigma+2\phi_1}\cosh\phi_5\sinh\phi_2,\\
\phi_2'&=&2ge^\sigma\textrm{sech}\phi_5\left[(e^{2\phi_1}-1)\cosh\phi_2-\sinh\phi_2\right]\nonumber \\
& &+6a\kappa e^{-\sigma-2h}\textrm{sech}\phi_5\sinh\phi_2,\\
\phi_5'&=&-2ge^\sigma\sinh\phi_5\left[\cosh\phi_2-(e^{2\phi_1}-1)\sinh\phi_2\right]\nonumber \\
& &+6\kappa e^{-\sigma-2h}(b\cosh\phi_5-a\cosh\phi_2\sinh\phi_5),\\
\sigma'&=&\frac{3}{2}me^{-3\sigma}-\frac{1}{2}ge^\sigma\cosh\phi_5\left[\cosh\phi_2-(e^{2\phi_1}-1)\sinh\phi_2\right]\nonumber \\
& &+\frac{3}{2}\kappa e^{-\sigma-2h}(a\cosh\phi_2\cosh\phi_5-b\sinh\phi_5)+\frac{9(a^2-b^2)e^{\sigma-4h}}{8m},\\
h'&=&\frac{1}{2}ge^\sigma\cosh\phi_5\left[\cosh\phi_2-(e^{2\phi_1}-1)\sinh\phi_2\right]+\frac{1}{2}me^{-3\sigma}\nonumber \\
& &+\frac{3}{2}\kappa e^{-\sigma-2h}(a\cosh\phi_2\cosh\phi_5-b\sinh\phi_5)-\frac{9(a^2-b^2)e^{\sigma-4h}}{8m},\\
f'&=&\frac{1}{2}ge^\sigma\cosh\phi_5\left[\cosh\phi_2-(e^{2\phi_1}-1)\sinh\phi_2\right]+\frac{1}{2}me^{-3\sigma}\nonumber \\
& &-\frac{3}{2}\kappa e^{-\sigma-2h}(a\cosh\phi_2\cosh\phi_5-b\sinh\phi_5)+\frac{27(a^2-b^2)e^{\sigma-4h}}{8m} \, .
\end{eqnarray}
We now look at possible $AdS_2\times \mc{M}_{\textrm{K}4}$ fixed points. As in the previous section, setting $\phi_1'=\phi_2'=0$ in the first two BPS equations implies that $\phi_1=\phi_2=0$. By solving the remaining equations, we find two $AdS_2\times \mc{M}_{\textrm{K}4}$ solutions:
\begin{itemize}
\item $AdS_2$ critical point III:
\begin{eqnarray}
\phi_1&=&\phi_2=\phi_5=0,\qquad b=0,\qquad \sigma= \frac{1}{4}\ln\left[\frac{2m}{g}\right],\nonumber \\ 
h&=&\frac{1}{2}\ln\left[-\frac{3a\kappa}{\sqrt{2mg}}\right],\qquad
\ell=\frac{1}{2(2g^3m)^{\frac{1}{4}}}\, . \label{AdS2_III}
\end{eqnarray}
\item $AdS_2$ critical point IV:
\begin{eqnarray}
\phi_1&=&\phi_2=0,\qquad \phi_5=\varphi_0,\nonumber \\
h&=&\frac{1}{2}\ln\left[\frac{3\kappa e^{-2\sigma}(a+b-ae^{2\varphi_0}+be^{2\varphi_0})}{g(e^{2\varphi_0}-1)}\right],\nonumber \\
\sigma&=&-\frac{1}{4}\ln\left[\frac{ge^{-\varphi_0}(e^{2\varphi_0}-1)(a+b+ae^{2\varphi_0}-be^{2\varphi_0})}{4m[a(e^{2\varphi_0}-1)-b(1+e^{2\varphi_0})]}\right],\nonumber \\
\ell&=&\frac{\sqrt{e^{2\varphi_0}-1}}{(1+e^{2\varphi_0})\left[4g^3m(1+e^{2\varphi_0}+\sqrt{3-2e^{2\varphi_0}+3e^{4\varphi_0}})\right]^{\frac{1}{4}}}
\label{AdS2_IV}
\end{eqnarray}
with $\varphi_0$ being a solution of the following equation
\begin{equation}
b^2\sinh3\varphi_0+4a^2\sinh^3\varphi_0+9b^2\sinh\varphi_0-8ab\cosh^3\varphi_0=0\, .
\end{equation}    
\end{itemize}
As in the case of $AdS_2$ critical point II, the explicit form of $\varphi_0$ is very complicated, so we will not give it here.
\\
\indent  For both of these critical points, $AdS_2\times \mc{M}_{\textrm{K}4}$ solutions exist only for $\kappa=-1$ leading to black holes with $\mc{M}_{\textrm{K}4}^-$ horizons. Examples of solutions interpolating between the $AdS_6$ vacuum and $AdS_2\times \mc{M}^-_{\textrm{K}4}$ critical point III with different values of $m$ are shown in figure \ref{fig3}. In these solutions, we have set $g=3m$, $b=0$ and $\phi_1=\phi_2=0$. The red, green and blue lines respectively represent solutions with $m=1,2,3$. The black hole entropy in this case is given by
\begin{eqnarray}
S_{\textrm{BH}}&=&\frac{1}{4G_{\textrm{N}}}e^{4h}\textrm{vol}(\mc{M}_4),\nonumber \\
&=&\frac{9a^2\textrm{vol}(\mc{M}^-_{\textrm{K}4})}{8gmG_{\textrm{N}}}\, .
\end{eqnarray}  
 \indent Similarly, examples of black hole solutions with the near horizon geometry given by critical point IV are shown in figure \ref{fig4} for $\phi_1=\phi_2=0$, $g=3m$ and $m=1$. The green, purple and orange lines respectively represent solutions with $b=-\frac{1}{8},-\frac{1}{10},-\frac{1}{12}$. In this case, the entropy of the black hole is given by
\begin{equation}
S_{\textrm{BH}}=\frac{9\left[(a^2+b^2)\sinh2\varphi_0-ab\cosh2\varphi_0\right]\textrm{vol}(\mc{M}^-_{\textrm{K}4})}{16gmG_{\textrm{N}}\sinh\varphi_0}\, .
\end{equation}  

\subsection{Black holes with Cayley four-cycle horizon}
\begin{figure}
         \centering
               \begin{subfigure}[b]{0.4\textwidth}
                 \includegraphics[width=\textwidth]{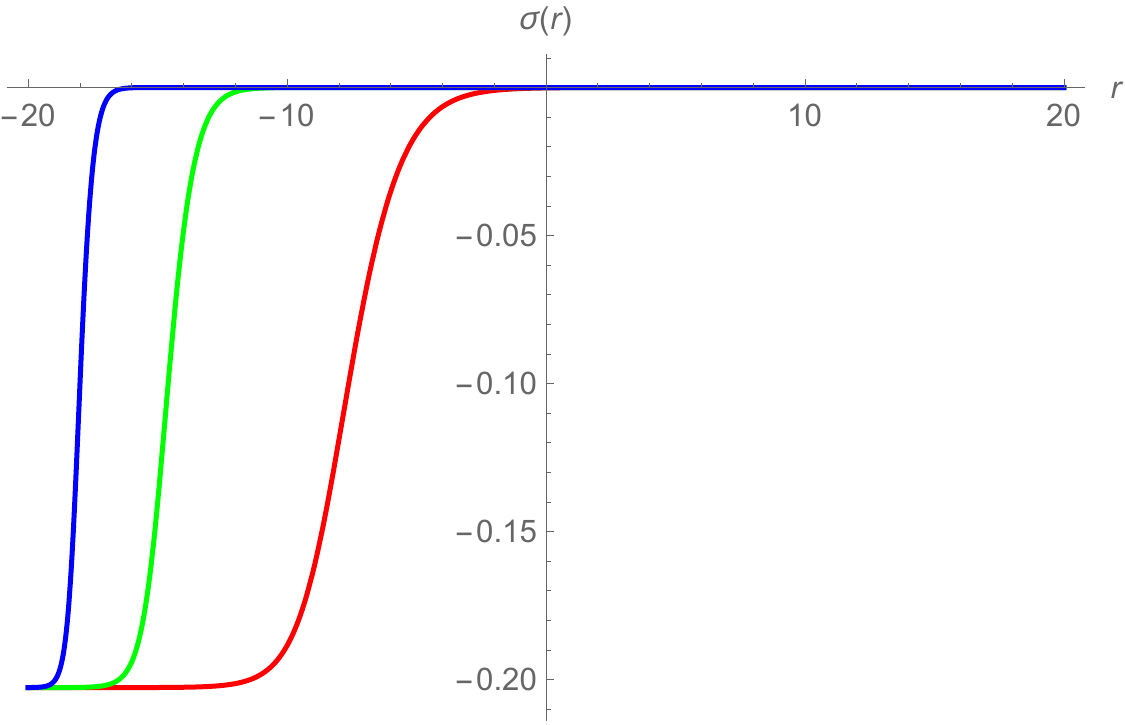}
                 \caption{Solutions for $\sigma(r)$}
         \end{subfigure}
          \begin{subfigure}[b]{0.4\textwidth}
                 \includegraphics[width=\textwidth]{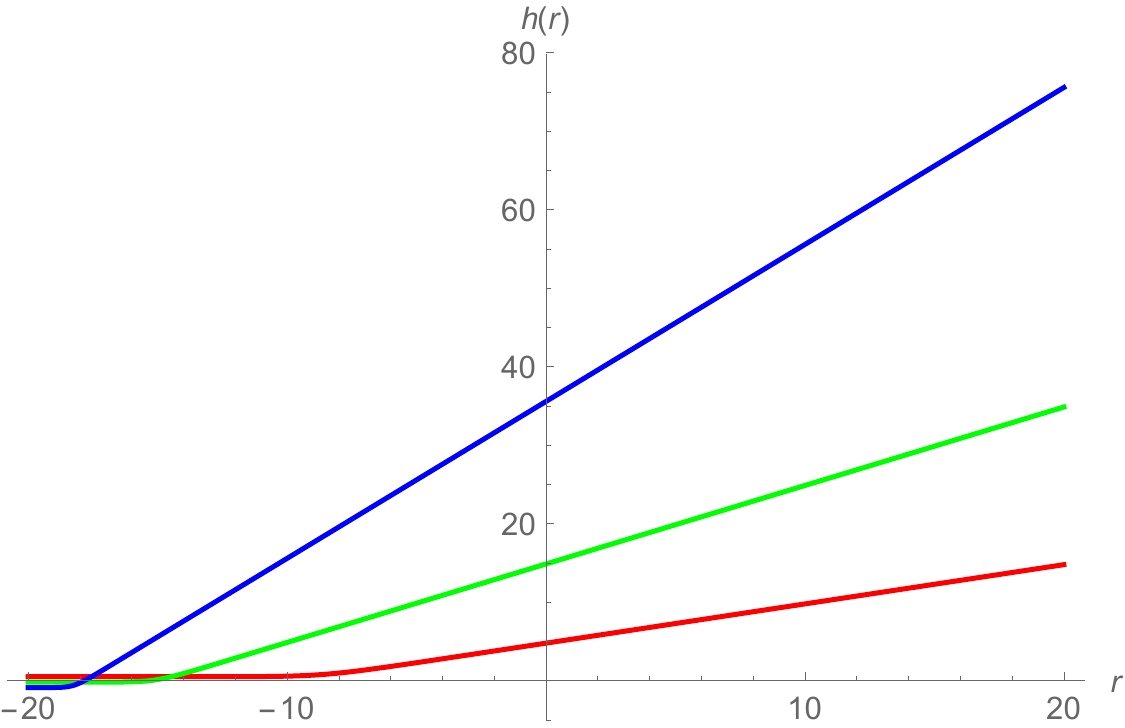}
                 \caption{Solutions for $h(r)$}
         \end{subfigure}\\
          \begin{subfigure}[b]{0.4\textwidth}
                 \includegraphics[width=\textwidth]{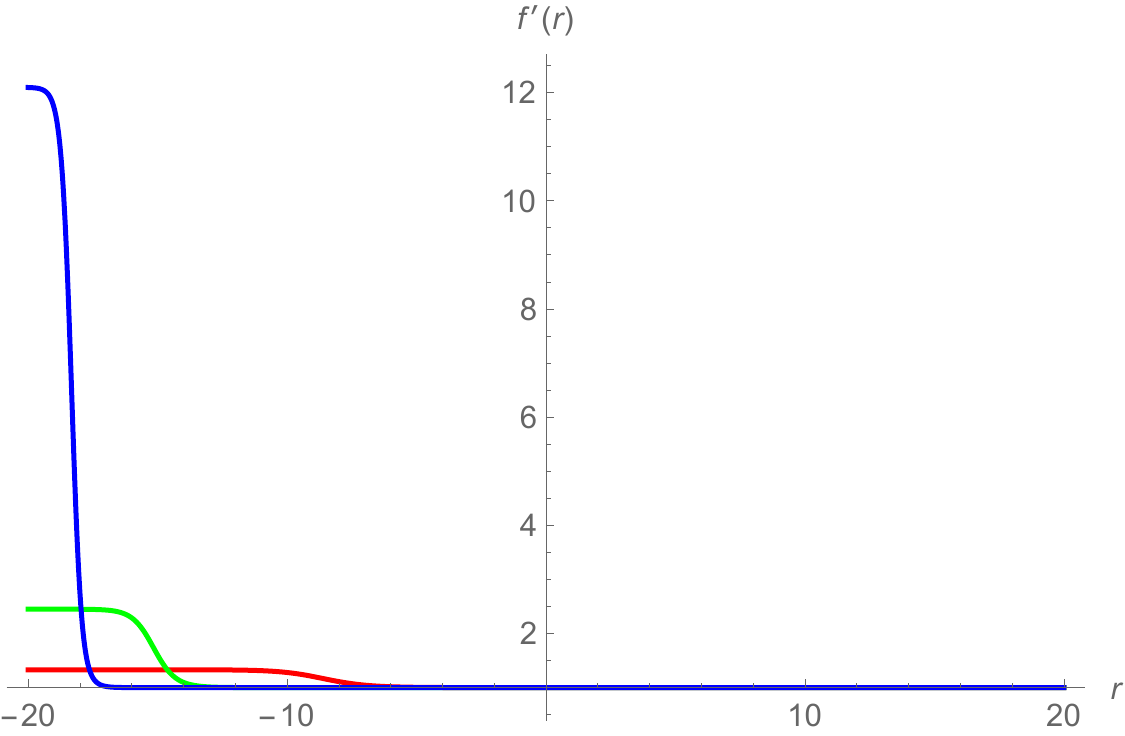}
                 \caption{Solutions for $f'(r)$}
         \end{subfigure}
\caption{Supersymmetric $AdS_6$ black holes interpolating between the $AdS_6$ vacuum and the near horizon geometry $AdS_2\times \mc{M}_{\textrm{C}4}^-$ (critical point V) for $m=\frac{1}{4}$ (red), $m=\frac{1}{2}$ (green), $m=1$ (blue).}\label{fig5}
 \end{figure}

In this section, we consider black hole solutions with the horizon given by a Cayley four-cycle $\mc{M}_{\textrm{C}4}$ with the metric ansatz
\begin{equation}
ds^2=-e^{2f(r)}dt^2+dr^2+e^{2h(r)}ds^2_{\mc{M}_{\textrm{C}4}}\, .
\end{equation}
The metric on $\mc{M}_{\textrm{C}4}$ can be written as
\begin{equation}
ds^2_{\mc{M}_{\textrm{C}4}}=d\rho^2+F_\kappa(\rho)^2(\tau_1^2+\tau_2^2+\tau_3^2)
\end{equation}
in which $\tau_i$ are $SU(2)$ left-invariant one-forms as in the case of Kahler four-cycle and $F_\kappa(\rho)$ defined as in \eqref{F_def}.
\\
\indent In this case, the isometry of the four-cycle is given by $SO(4)\sim SU(2)_+\times SU(2)_-$. We will perform a twist by turning on $SO(3)$ gauge fields and identify this $SO(3)$ with the self-dual part $SU(2)_+$ of the isometry. With the vielbein chosen as
\begin{equation}
e^{\hat{t}}=e^fdt,\qquad e^{\hat{r}}=dr,\qquad e^{\hat{i}}=e^hF_\kappa(\rho)\tau_i,\qquad e^{\hat{4}}=e^hd\rho,
\end{equation}
non-vanishing components of the spin connection are given as follows
\begin{eqnarray}
& &{\omega^{\hat{t}}}_{\hat{r}}=f'e^{\hat{r}},\qquad {\omega^{\hat{i}}}_{\hat{r}}=h'e^{\hat{4}},\qquad {\omega^{\hat{i}}}_{\hat{r}}=h'e^{\hat{i}},\qquad \hat{i}=1,2,3,\nonumber \\
& &{\omega^{\hat{i}}}_{\hat{4}}=e^{-h}\frac{F'_\kappa(\rho)}{F_\kappa(\rho)}e^{\hat{i}},\qquad {\omega^{\hat{i}}}_{\hat{j}}=\frac{e^{-h}}{F_\kappa(\rho)}\epsilon_{\hat{i}\hat{j}\hat{k}}e^{\hat{k}}\, .
\end{eqnarray}
As in \cite{ISO3_flow} and \cite{ISO3_defect}, there are two $SO(3)$ singlet scalars from $SO(4,4)/SO(4)\times SO(4)$ coset with the coset representative given by
\begin{equation}
L=e^{\phi \hat{Y}_1}e^{\varphi \hat{Y}_2}\, .\label{L_SO3}
\end{equation} 
The $SO(4,4)$ non-compact generators take the form of
\begin{equation}
\hat{Y}_1=Y_{11}+Y_{22}+Y_{33}\qquad \textrm{and}\qquad \hat{Y}_2=Y_{04}\, .
\end{equation}
The coset representative \eqref{L_SO3} leads to the composite connection
\begin{equation}
Q_{AB}=-\frac{i}{2}gA^r\sigma^r_{AB}
\end{equation}
in which the $SO(3)$ gauge fields are denoted by $A^r$. In order to cancel the internal spin connection on $\mc{M}_{\textrm{C}4}$, we choose the gauge fields as follows
\begin{equation}
A^r=a(F'_\kappa(\rho)+1)\delta^r_i\tau_i\, .
\end{equation}
By the same analysis as in the previous cases, we impose the following projectors 
\begin{equation}
\gamma_{\hat{i}\hat{4}}\epsilon_A=\frac{1}{2}\epsilon_{\hat{i}\hat{j}\hat{k}}\gamma_{\hat{j}\hat{k}}\epsilon_A=-i\delta^r_i\sigma^r_{AB}\epsilon^B\label{CY4_proj}
\end{equation}
and the twist condition
\begin{equation}
ga=-1\, .
\end{equation}
All the conditions from $\delta \psi_{\hat{i}A}$ and $\delta\psi_{\hat{4}A}$ then reduce to a single BPS equation for $h(r)$. We also note that there are only three independent projectors in \eqref{CY4_proj} leading to $\frac{1}{8}$-BPS $AdS_2\times \mc{M}_{\textrm{C}4}$ solutions with two supercharges. The full black hole solutions with non-constant scalars however are $\frac{1}{16}$-BPS due to the additional $\gamma^{\hat{r}}$ projector given in \eqref{gamma_r_proj}.
\\
\indent Before presenting the BPS equations, we note the explicit form of the gauge field strength tensors 
\begin{equation}
F^r=\kappa a\delta^r_ie^{-2h}\left(e^{\hat{i}}\wedge e^{\hat{4}}+\frac{1}{2}\epsilon_{\hat{i}\hat{j}\hat{k}}e^{\hat{j}}\wedge e^{\hat{k}}\right)
\end{equation} 
and the two-form field
\begin{equation}
B_{\hat{t}\hat{r}}=\frac{3}{8}\frac{\kappa^2e^{2\sigma-4h}}{m^2\mc{N}_{00}}(a^2-b^2).
\end{equation}
Furthermore, it turns out that the field equations require $\varphi=0$. With all these, the resulting BPS equations read
\begin{eqnarray}
\phi'&=&\frac{1}{2}e^{-\sigma-2h-\phi}(e^{2\phi}-1)(2ge^{2h+2\sigma+2\phi}+a\kappa),\label{C4_eq1}\\
\sigma'&=&\frac{1}{4}ge^{\sigma+\phi}(e^{2\phi}-3)+\frac{3}{2}me^{-3\sigma}-\frac{3}{4}\kappa a e^{-\sigma-2h}\cosh\phi+\frac{3a^2e^{\sigma-4h}}{32m},\\
h'&=&\frac{1}{2m}e^{-3\sigma}-\frac{1}{4}ge^{\sigma+\phi}(e^{2\phi}-3)-\frac{3}{4}a\kappa e^{-\sigma-2h}
\cosh\phi-\frac{3a^2e^{\sigma-4h}}{32m},\\
f'&=&\frac{1}{2m}e^{-3\sigma}-\frac{1}{4}ge^{\sigma+\phi}(e^{2\phi}-3)+\frac{3}{4}a\kappa e^{-\sigma-2h}
\cosh\phi+\frac{9a^2e^{\sigma-4h}}{32m}\, .
\end{eqnarray}
There is one $AdS_2\times \mc{M}_{\textrm{C}4}$ solution given by
\begin{itemize}
\item $AdS_2$ critical point V:
\begin{eqnarray}
\phi&=&0,\qquad \sigma=\frac{1}{4}\ln\left[\frac{4m}{3g}\right],\nonumber \\
h&=&\frac{1}{2}\ln\left[\frac{a\kappa}{2m}\sqrt{\frac{3m}{g}}\right],\qquad \ell =\frac{1}{2\sqrt{2}g}\left(\frac{3g}{m}\right)^{\frac{1}{4}}\, .\label{AdS2_CY}
\end{eqnarray}
\end{itemize}
This solution only exists for $\kappa=-1$ leading to an $AdS_2\times \mc{M}^-_{\textrm{C}4}$ solution. We also point out that another possible solution of $\phi'=0$ condition is obtained by setting the second parentheses in \eqref{C4_eq1} to zero and taking $\kappa=1$. However, this does not lead to real solutions for all the other fields. 
\\
\indent The corresponding black hole entropy is given by
\begin{equation}
S_{\textrm{BH}}=\frac{3a^2\textrm{vol}(\mc{M}_{\textrm{C}4}^-)}{16gmG_{\textrm{N}}},
\end{equation}
and some examples of numerical black hole solutions for $g=3m$ and different values of $m$ are shown in figure \ref{fig5}. The red, green and blue lines represent solutions with $m=\frac{1}{4},\frac{1}{2},1$, respectively.   

\section{Conclusions}\label{conclusion}
We have constructed a number of supersymmetric $AdS_6$ black hole solutions from six-dimensional $F(4)$ gauged supergravity coupled to four vector multiplets with $ISO(3)\times U(1)$ gauge group. These solutions interpolate between the supersymmetric $AdS_6$ vacuum and near horizon geometries of the form $AdS_2\times \mc{M}_4$. By performing topological twists using $SO(2)\times U(1)$ gauge fields, we have found black hole solutions with $AdS_2\times H^2\times H^2$, $AdS_2\times S^2\times H^2$ and $AdS_2\times H^2\times S^2$ near horizon geometries. By extending the analysis to the case of $\mc{M}_4$ being a Kahler four-cycle, we have also found two $AdS_2\times \mc{M}_{\textrm{K}4}^-$ solutions with $\mc{M}_{\textrm{K}4}^-$ denoting a negatively curved Kahler four-cycle. 
\\
\indent Moreover, by performing a twist using $SO(3)\subset ISO(3)$ gauge fields, we have found a black hole solution with $AdS_2\times \mc{M}_{\textrm{C}4}^-$ near horizon geometry for $\mc{M}^-_{\textrm{C}4}$ being a negatively curved Cayley four-cycle. In all cases, we have also given numerical black hole solutions interpolating between these near horizon geometries and the $AdS_6$ vacuum. On the other hand, the solutions can be interpreted as holographic RG flows across dimensions from five-dimensional $N=2$ SCFT dual to the $AdS_6$ vacuum to superconformal quantum mechanics dual to the $AdS_2$ geometry. Since the $ISO(3)\times U(1)$ matter-coupled $F(4)$ gauged supergravity can possibly be obtained from a consistent truncation of type IIB theory on $S^2\times \Sigma$, the solutions found here could give rise to new $AdS_2\times \mc{M}_4\times S^2\times \Sigma$ solutions in type IIB theory with $\mc{M}_4$ being $H^2\times H^2$, $S^2\times H^2$, $H^2\times S^2$, $\mc{M}_{\textrm{K}4}^-$ and $\mc{M}_{\textrm{C}4}^-$.  
\\
\indent It would be interesting to compute the black hole entropy for the solutions given here from the topologically twisted indices of the dual five-dimensional $N=2$ SCFT using the results of supersymmetric localization as in \cite{5D_twist_index1,5D_twist_index2,5D_twist_index3}. Uplifting the black hole solutions found in this paper to type IIB theory and identify possible brane configurations with near horizon geometries given by the $AdS_2\times \mc{M}_4\times S^2\times \Sigma$ solutions mentioned above is also of particular interest. However, as pointed out in \cite{Henning_Malek_AdS7_6}, it is currently unclear whether there are globally regular supersymmetric $AdS_6$ solutions of type IIB supergravity obtained from uplifting the solutions of $ISO(3)\times U(1)$ six-dimensional gauged supergravity. This is due to the fact that in order to obtain a complete truncation ansatz, there should exist a solution to a set of constraints given in \cite{Henning_Malek_AdS7_6}. Therefore, at this stage, it is also not certain whether there is a five-dimensional field theory dual to the solutions obtained in this paper. Finally, constructing $AdS_6$ black hole solutions with horizon geometries given by four-dimensional orbifolds as in the recent results \cite{spindle_BH1,spindle_BH2,4_orbifold_BH1,4_orbifold_BH2,4_orbifold_BH3} is certainly worth considering. We leave these issues for future works.             
\vspace{0.5cm}\\
{\large{\textbf{Acknowledgement}}} \\
This work is funded by National Research Council of Thailand (NRCT) and Chulalongkorn University under grant N42A650263. 

\end{document}